\documentclass[twocolumn,preprintnumbers,prx]{revtex4-2}
\usepackage{graphicx}
\graphicspath{{figure/}}
\usepackage{dcolumn}
\usepackage{physics} 
\usepackage{hyperref}
\hypersetup{
    colorlinks=true,
    linkcolor=blue,
    filecolor=blue,      
    urlcolor=blue,
    pdftitle={Overleaf Example},
    pdfpagemode=FullScreen,
    }
    
\usepackage[english]{babel}
\usepackage[autostyle]{csquotes}
\usepackage{listings}
\usepackage{lstautogobble} 
\usepackage{amsmath,amsfonts,amsthm,amssymb,bm} 
\usepackage[capitalize]{cleveref}
\usepackage{indentfirst}
\usepackage[utf8]{inputenc}
\usepackage{geometry}
\usepackage{bm}
\usepackage{MnSymbol}
\usepackage{mathrsfs}

\usepackage[font=footnotesize,labelfont=bf,margin=1cm,justification=justified,format=plain]{caption}
\usepackage{subcaption}

\begin{document}

\title{Non-Fermi liquid behavior in a simple model of Fermi arcs and pseudogap}
\author{Ruojun Wang and Kun Yang}
\affiliation{Department of Physics and National High Magnetic Field Laboratory,
Florida State University, Tallahassee, Florida 32306, USA}

\begin{abstract}

We consider a perturbed version of a very simple and exactly solvable model that supports Fermi arcs and pseudogap in its ground state and excitation spectrum, which includes Hubbard-like interactions in both momentum and real spaces. We find the combined effects give rise to non-Fermi liquid behavior in the electron self-energy. Comparison will be made with phenomenology of high temperature cuprate superconductors.

\end{abstract}

\date{\today}

\maketitle

\section{Introduction}

The normal state of high transition temperature ($T_c$) cuprate superconductors are known to exhibit non-Fermi liquid behavior in the underdoped regime.
Among their mysterious properties \cite{Proust}, they support pseudogaps and Fermi arcs \cite{Yoshida} instead of closed Fermi surfaces of ordinary Fermi liquid \cite{Book}. Non-Fermi liquid behavior is also manifested in the lack of coherence in quasiparticle excitations and unusual transport properties \cite{Proust,Yoshida}. Understanding such non-Fermi liquid physics is an exciting challenge we face.

In an earlier paper \cite{Yang21} one of us introduced an extremely simple and exactly solvable model, and showed that Fermi arcs and pseudogap appear very naturally (and hand-in-hand) in its ground state and excitation spectrum. That model is a variant of a model introduced by Hatsugai and Kohmoto (HK) \cite{hk} (a model similar to that of HK was considered earlier by Baskaran \cite{Baskaran}).
An unusual property of this model \cite{Yang21} (which we refer to as HKY model from now on) is that all quasiparticle and quasihole excitations are sharp, albeit being gapped in the pseudogap region. This is, of course, opposite to non-Fermi liquids where quasiparticle and quasihole excitations are incoherent, rendering the electron spectral functions very broad. The sharpness of the electron spectral function in the HKY model is the consequence of the fact that its interaction is local in momentum space and only gives rise to forward scattering. To remove this artifact, in the present paper we perturb the HKY model with a (real space) Hubbard interaction, and calculate its contribution to electron self-energy. We demonstrate that the combined effects of the Hubbard and HKY interactions render the quasiparticle and quasihole excitations incoherent, consistent with the cuprate phenomenology.

The rest of the paper is organized as what follows. In Sec. \ref{sec: model} we introduce the HKY model perturbed by the Hubbard interaction, and its meanfield solution which gives rise to Fermi arcs and pseudogap regions. In Sec. \ref{sec: Feynman rules} we set up the Feynman rules for perturbative treatments of interactions, and demonstrate that all non-vanishing diagrams involving HKY interactions form particle-particle and particle-hole ladders that can be summed exactly. In Sec. \ref{Sec: Self-Energy} we calculate the electron self-energy to the 2nd order in Hubbard interaction, and demonstrate its imaginary part remains finite in the low-energy limit, resulting in non-Fermi liquid behavior. A brief summary is provided in Sec. \ref{sec: Summary}.

\section{Model and mean-field solution}
\label{sec: model}

We start by considering the HKY model:
\begin{equation}
H_{\text{HKY}}=\sum_{{\bf k}}[\epsilon_{{\bf k}}(\hat{n}_{{\bf k}\uparrow} + \hat{n}_{{\bf k}\downarrow})+u_{{\bf k}} \hat{n}_{{\bf k}\uparrow}\hat{n}_{{\bf k}\downarrow}],
\label{eq:HKY}
\end{equation}
where $\epsilon_{\vb{k} }$ is the single particle energy,  $n_{\vb{k}\sigma} = c_{\vb{k}\sigma }^\dagger c_{\vb{k}\sigma} $ is the fermion occupation for momentum $\vb{k}$ and spin $\sigma = \uparrow, \downarrow$, and $u_{\vb{k} } $ is the interaction energy between the spin-up and spin-down particles. When $u_{\vb{k} }$ is a constant, the model is reduced to the HK model \cite{hk}. 

While (\ref{eq:HKY}) is exactly solvable, in preparation for the breakdown of solvability once the (real space) Hubbard (or any other generic) interaction is introduced we first introduce a meanfield solution to (\ref{eq:HKY}), which we will use as the starting point for perturbation theory later on. Note this meanfield solution is exact for the ground state and single particle/hole excitations. We separate the operator $\hat{n}_{\vb{k}\sigma}$ into its expectation value and fluctuation:
\begin{eqnarray} \label{eq:}
    \hat{n}_{\vb{k}\sigma} = 
    {n}_{\vb{k}\sigma} + \delta \hat{n}_{\vb{k}\sigma}
    , 
\end{eqnarray}
where $n_{\vb{k}\sigma} = \expval{\hat{n}_{\vb{k}\sigma}} $, and write the Hamiltonian as: 
\begin{eqnarray} \label{eq:Heff}
    H_{\text{HKY}} = H_0 + H_1 + H_2 ,
\end{eqnarray}
such that  
\begin{eqnarray} \label{eq:}
& \label{eq:H0}
H_0 = 
\sum_{\vb{k} \sigma } 
    E_{\vb{k} \sigma } 
    \hat{n}_{\vb{ k } \sigma } 
, \\& \label{eq:H1}
H_1 =  
-\sum_{\vb{k } \sigma } 
n_{\vb{ k }, -\sigma } u_{\vb{ k } }  
\hat{n}_{\vb{ k } \sigma } 
, \\& \label{eq:H2}
H_2 = 
\sum_{\vb{k } } u_{\vb{ k } } \hat{{ n }}_{\vb{ k } \uparrow }  \hat{{ n }}_{\vb{ k } \downarrow } 
, 
\end{eqnarray}
where $E_{\bf k \sigma } = \epsilon_{\bf k } + n_{\vb{k}, -\sigma} u_{\bf k} $, is the single-particle energy within the Hartree approximation and we denote $-\sigma$ as the opposite spin of $\sigma$. The ground state of $H_0$, which is also the {\em exact} ground state of $H_{\text{HKY}}$, has the following occupation pattern:
\begin{eqnarray}
n_{{\bf k}}= \left\lbrace\begin{array}{cc}
0, &\epsilon_{{\bf k}} > 0 \hskip 0.2 cm {\rm and} \hskip 0.2 cm \epsilon_{{\bf k}} + u_{{\bf k}} > 0\\
1, &\epsilon_{{\bf k}} < 0 \hskip 0.2 cm {\rm and} \hskip 0.2 cm \epsilon_{{\bf k}} + u_{{\bf k}} > 0\\
2, &\epsilon_{{\bf k}} < 0 \hskip 0.2 cm {\rm and} \hskip 0.2 cm \epsilon_{{\bf k}} + u_{{\bf k}} < 0 \end{array}\right. 
\label{eq:occupation}
\end{eqnarray}
and those regions are distinguished by the surfaces defined by 
\begin{eqnarray} \label{eq:}
    \label{eq:boundary2}
    &&\epsilon_{{\bf k}} = 0,\\
    \label{eq:boundary1}
    &&\epsilon_{{\bf k}} + u_{{\bf k}} = 0,\\
    \label{eq:u_k}
    && u_{{\bf k}} = 0.
\label{eq:boundaries}
\end{eqnarray}

As pointed out in \cite{Yang21}, we have pseudo-Fermi surfaces across which occupation numbers change by 2 ($\Delta n_{\vb{k}} = 2 $) where there is a single-particle energy gap of $|u_{\vb{k}}|/2$ (i.e., pseudogap), and Fermi-arcs across which occupation numbers change by 1 ($\Delta n_{\vb{k}} = 1 $) with no such gap. In the region with $n_{\vb{k}}= 1$, each state can be occupied by either spin-up or spin-down fermions, resulting in a massive degeneracy. In order to remove this degeneracy, we can introduce an infinitesimal Zeeman splitting $\Delta_Z$ between the spin-up and spin-down fermions so that the $n_{\vb{k}}=1$ regions are occupied by the spin-down fermions only in the ground state. The Fermi arcs are then the Fermi surfaces for spin-up and spin-down fermions respectively, albeit they are not closed (hence arcs).
The single-particle Green's function of $H_0$, again the same as the exact Green's function of $H_{\text{HKY}}$, is 
\begin{equation} \label{eq:gf0}
\begin{split}
    G_{ \sigma}^0 ( \omega\vb{k} ) 
=&
\frac{ 
	( 1 - n_{\vb{ k } \sigma } ) ( 1- n_{\vb{ k }, -\sigma } )
}
{ 
	\omega 
	+ i \eta 
	- \hbar^{ -1 }
	\epsilon_{\vb{ k } }   
}  
+ 
\frac{ 
	( 1 - n_{\vb{ k } \sigma } ) n_{ \vb{ k }, -\sigma }  
}
{ 
	\omega 
	+ i \eta 
	- \hbar^{ -1 }
	\left( 
	\epsilon_{\vb{k } } 
	+ u_{\vb{ k } }
	\right) 
}
\\&
+ 
\frac{
	n_{ \vb{ k } \sigma } ( 1- n_{ \vb{ k }, -\sigma } )
}{
	\omega 
	- i \eta 
	- 
	\hbar^{ -1 }
	\epsilon_{\vb{k } } 
}
+ 
\frac{
	n_{ \vb{ k } \sigma } n_{ \vb{k }, -\sigma }
}{
	\omega 
	- i \eta 
	- 
	\hbar^{ -1 }
	\left( 
	\epsilon_{\vb{k } } + u_{\vb{k } }  
	\right) 
}
\\=&
\frac{ 
	( 1 - n_{\vb{ k } \sigma } )
}
{ 
	\omega 
	+ i \eta 
	- \hbar^{ -1 }
	E_{\vb{ k }\sigma }   
}  
+ 
\frac{
	n_{ \vb{ k } \sigma } 
}{
	\omega 
	- i \eta 
	- 
	\hbar^{ -1 }
	E_{\vb{k }, -\sigma} 
}
,
\end{split}
\end{equation}
where $\eta$ is an infinitesimal positive. 

In addition to the Hamiltonian in Eq. (\ref{eq:Heff}), we consider a perturbing Hubbard interaction: 
\begin{align} \label{eq:hubbard}
    H_{\text{Hubbard}} 
    =&
    V \sum_i \hat{n}_{i\uparrow}\hat{n}_{i\downarrow}
    \\=&
    \frac{V}{N} \sum_{\vb{k}\vb{k}' \vb{q} }
        c_{\vb{k}+\vb{q} \uparrow }^\dagger 
        c_{\vb{k}'-\vb{q} \downarrow }^\dagger
        c_{\vb{k}' \downarrow }
        c_{\vb{k} \uparrow}
    , 
\end{align}
where $V$ is the interaction strength, $i$ is site index and $N$ is system size. Hence, the full Hamiltonian we would like to consider is the sum of the HKY Hamiltonian and the Hubbard interaction, 
\begin{eqnarray} \label{eq:H}
    H = H_{\text{HKY} } + H_{\text{Hubbard}} = H_0 + H', 
\end{eqnarray}
where we treat $H_1$, $H_2$, and $H_{\text{Hubbard}}$ perturbatively such that
\begin{eqnarray} \label{eq:H'}
    H' = H_1 + H_2 + H_{\text{Hubbard}}
    . 
\end{eqnarray}

\begin{figure}[!h]
    \centering
\begin{subfigure}[t]{0.12\textwidth} 
    \centering
    \includegraphics[width=1.5cm]{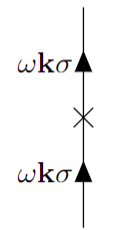}
    \caption{}
    \label{fig:v_mf}
\end{subfigure}
\begin{subfigure}[t]{0.12\textwidth}
    \centering
    \includegraphics[width=5cm]{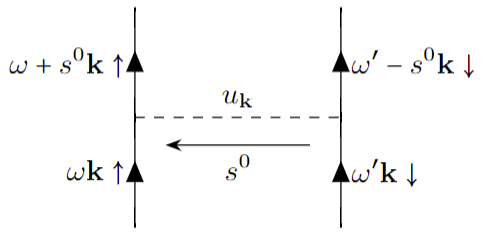}
    \caption{}
    \label{fig:v_hk}
  \end{subfigure}
\\
    \begin{subfigure}[t]{0.2\textwidth} 
    \centering
    \includegraphics[width=5cm]{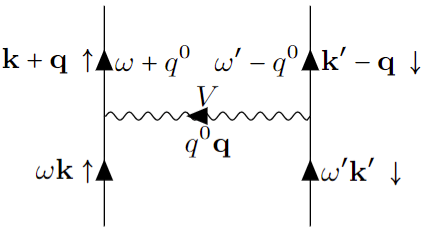}
    \caption{}
    \label{fig:v_hb}
  \end{subfigure}
    \caption{Vertices given by the interaction resulting from (a) $H_1$, (b) $H_2$, and (c) $H_{\text{Hubbard}}$. }
    \label{fig:vertices}
\end{figure}

\section{Feynman rules and Ladder Sums}

\label{sec: Feynman rules}
\begin{figure}
\center
\includegraphics[width=6cm]{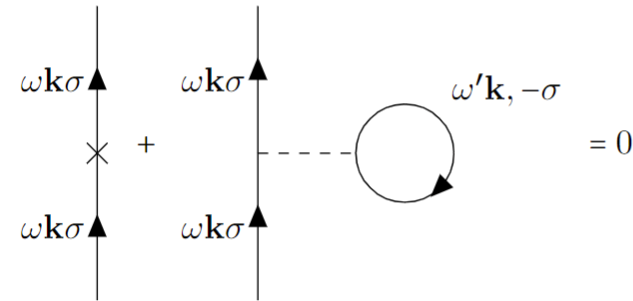}
\caption{An example of diagrams that are not part of the particle-particle or particle-hole ladder of HKY interaction ($H_2$). All such diagrams vanish.}
\label{fig:feynman_sum}
\end{figure}

The Feynman rules can be established following standard text books \cite{FW}. We introduce a solid line to denote the unperturbed Green's function $G_\sigma^0 (\omega, \vb{k}) $ in Eq. (\ref{eq:gf0}), a cross symbol to denote the interaction given by $H_1$ in Eq. (\ref{eq:H1}), a dashed line to denote the one given by $H_2$ in Eq. (\ref{eq:H2}), and a wavy line to denote the Hubbard interaction in Eq. (\ref{eq:hubbard}). Hence, it is necessary to consider those three kinds of vertices in Fig. \ref{fig:vertices}. In terms of these diagrammatic components, we find the diagrams in Fig. \ref{fig:feynman_sum} cancel each other, where the second one is the Hartree diagram in terms of the $H_2$ interaction. 
Note this cancellation occurs not only when they stand alone in the first order diagram illustrated here, but also when they are embedded in higher order diagrams. This cancellation, guaranteed by the self-consistent Hartree condition, is a significant simplification, as we can now drop all diagrams that involve the cross symbol given by Fig. \ref{fig:v_mf} and/or a Hartree bubble. Other than this simplification, the Feynman rules are the same as the usual ones \cite{FW}.

\begin{figure}
\center
\includegraphics[width=3cm]{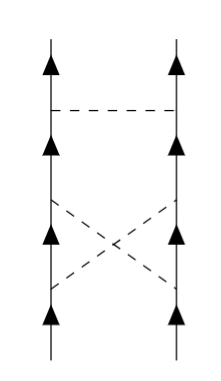}
\caption{An example of diagrams that are not part of the particle-particle or particle-hole ladder of HKY interaction ($H_2$). All such diagrams vanish.}
\label{fig:mix}
\end{figure}

\begin{widetext} 
In addition to Fig. \ref{fig:feynman_sum}, another major simplification is all other contributions from $H_2$ interaction can be organized as particle-particle and particle-hole ladder diagrams illustrated in Fig. \ref{fig:ladder_pp} and Fig. \ref{fig:ladder_ph}. 
\begin{figure}
\center
\includegraphics[width=15cm]{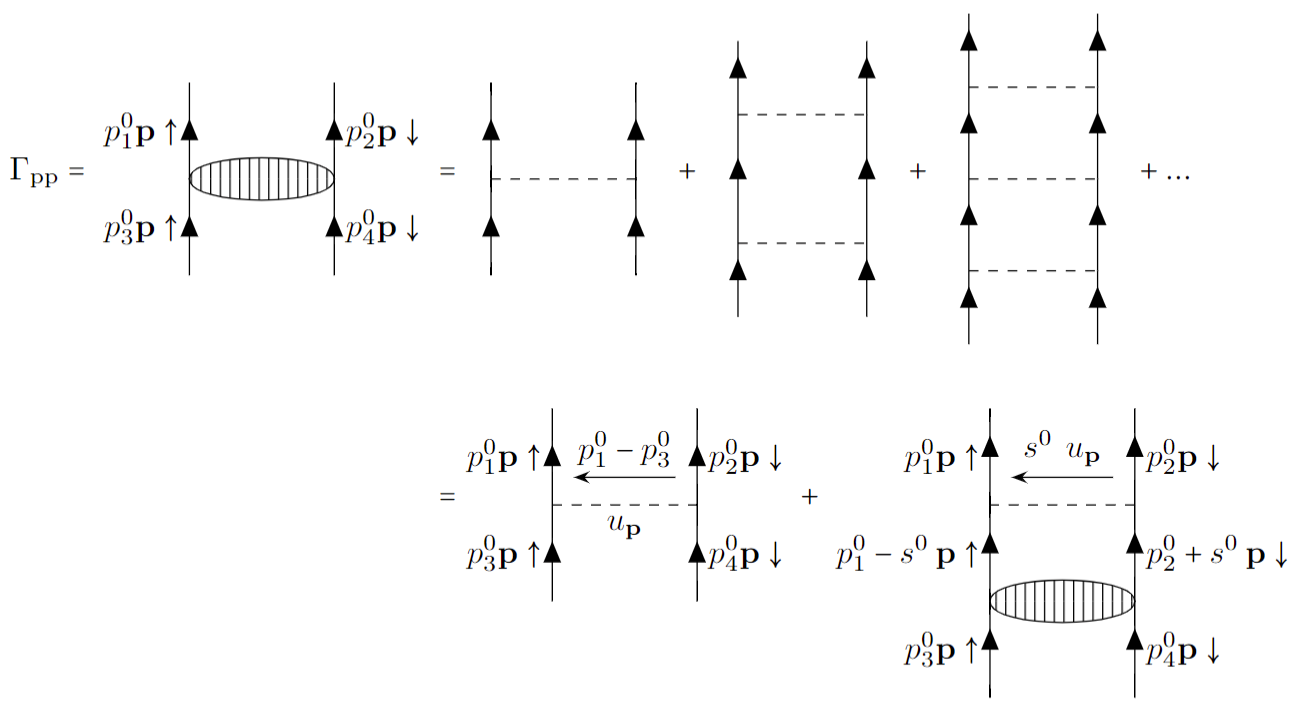}
\caption{Particle-particle ladder diagrams.}
\label{fig:ladder_pp}
\end{figure}
\begin{figure}
\center
\includegraphics[width=15cm]{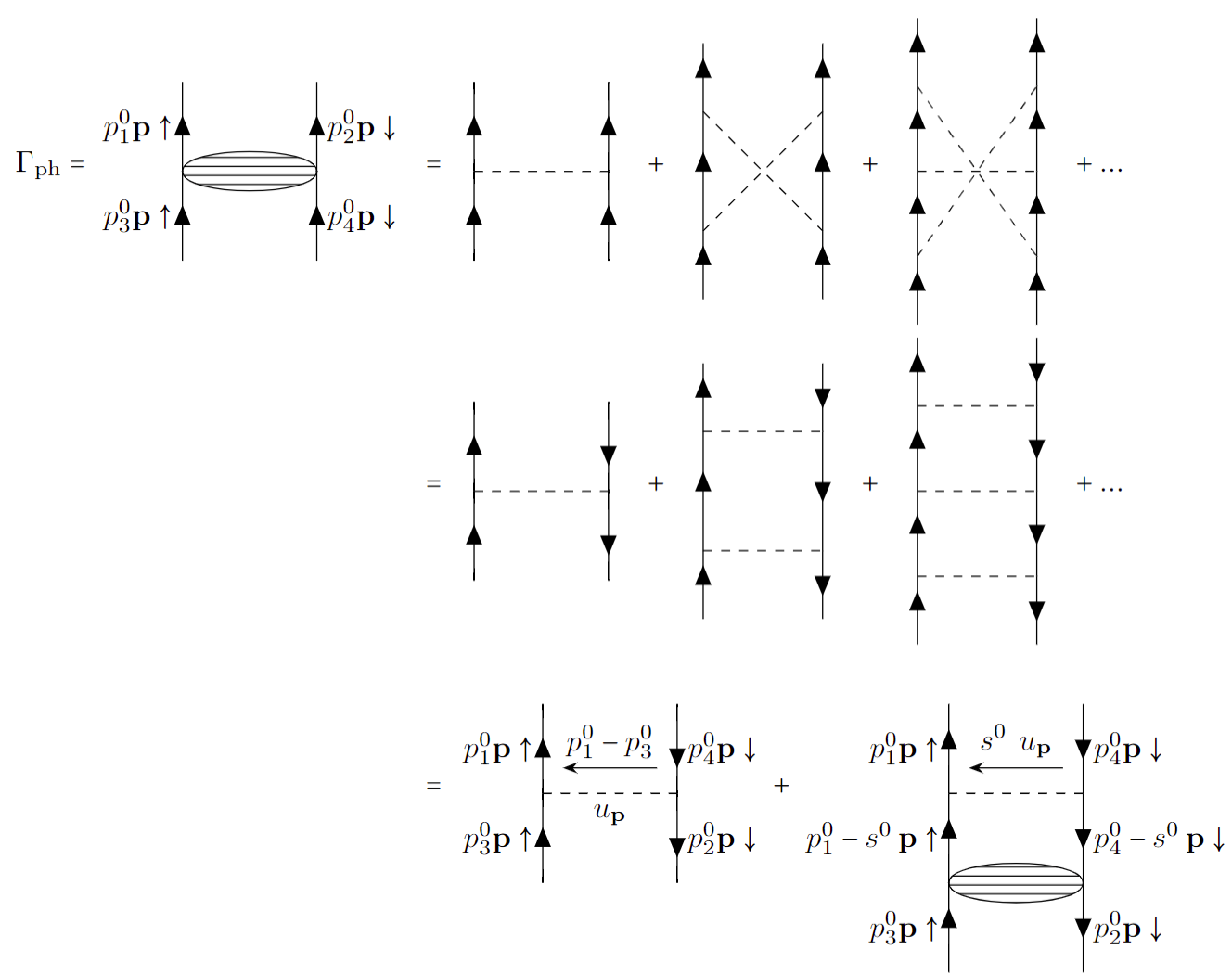}
\caption{Particle-hole ladder diagrams.}
\label{fig:ladder_ph}
\end{figure}
The first line of the equation in Fig. \ref{fig:ladder_ph} includes all particle-particle crossing diagrams, which are equivalent to the particle-hole ladder diagrams on the following line. Fig. \ref{fig:ladder_pp} and \ref{fig:ladder_ph} are the only nonzero contributions given by the $H_2$ term. This is because $H_2$ only gives rise to forward scattering, and cannot create particle-hole pairs. Also for this reason these ladder diagrams can be summed up easily because they form geometric series. To see this we inspect the corresponding Bethe-Salpeter equations \cite{FW}
\begin{equation}	
\begin{split}
	\Gamma_{\text{pp}} & (p_1^0 + p_2^0, \vb{ p } ) =  
	u (\vb{ p } ) +
	u (\vb{p} )
	\Gamma (p_1^0 + p_2^0, \vb{ p } ) 
	\int \frac{d q^0 }{2\pi } 
	G_\sigma^0 
	    \left(
	        \frac{p_1^0 + p_2^0}{2} + q^0, 
	        \vb{ p } 
	   \right) 
	G_{-\sigma }^0 
	    \left(
	        \frac{p_1^0 + p_2^0}{2} - q^0, 
	        \vb{ p } 
	   \right) 
	, 
\end{split}
\end{equation}
and 
\begin{equation}	
\begin{split}
	\Gamma_{\text{ph}} & (p_1^0-p_4^0, \vb{ p } ) =  
	u (\vb{ p } ) +
	u (\vb{p} )
	\Gamma (p_1^0-p_4^0, \vb{ p } ) 
	\int \frac{d q^0 }{2\pi } 
	G_\sigma^0 
	    \left(
	        \frac{p_1^0-p_4^0}{2} + q^0, 
	        \vb{ p } 
	   \right) 
	G_{-\sigma }^0 
	    \left(
	        q^0 - \frac{p_1^0-p_4^0}{2}, 
	        \vb{ p } 
	   \right) 
	, 
\end{split}
\end{equation}
where $p_1^0$, $p_2^0$ and $p_4^0$ are the frequencies carried by the external propagators. Due to the forward-scattering nature there is no momentum integral, as a result $\Gamma$ does not enter the integrals on the RHS, allowing the integrals to be carried out explicitly, yielding
\begin{eqnarray} \label{eq:pp}
\begin{split}
\Gamma_{\text{pp} } (p_1^0-p_2^0, \vb{p} ) 
    =& 
\frac{ 
    u(\vb{p}) 
    (1-n_{\vb{p }\sigma }) 
    (1-n_{\vb{p }, -\sigma } ) 
}{ 
    1
    - \frac{ u(\vb{p}) }{
         \hbar ( p_1^0+p_2^0 ) 
         -
            \left( 
                E_{\vb{p }\sigma } + E_{\vb{p }, -\sigma } 
            \right)
         + i\eta 
      } 
} 
+
\frac{ 
   u(\vb{p})
   n_{\vb{p }\sigma } 
   n_{\vb{p }, -\sigma }
}{
    1
    + \frac{ u(\vb{p}) }{ 
        \hbar ( p_1^0+p_2^0 ) 
        - 
        \left( 
            E_{\vb{p }\sigma } + E_{\vb{p }, -\sigma } 
        \right)
         - i\eta 
    } 
} 
, 
\end{split}
\end{eqnarray}
\begin{eqnarray} \label{eq:ph}
\begin{split}
\Gamma_{\text{ph} } (p_1^0-p_4^0, \vb{p} ) 
    =& 
\frac{ u (\vb{p})
    (1-n_{\vb{p }\sigma })
    n_{\vb{p }, -\sigma }  
}{ 
    1
    - \frac{ u (\vb{p}) }{
         \hbar ( p_1^0 - p_4^0 ) 
         -
            \left( 
                E_{\vb{p }\sigma } - E_{\vb{p }, -\sigma } 
            \right)
         + i\eta 
 } 
} 
+
\frac{ u (\vb{p})
    n_{\vb{p }\sigma }
    (1-n_{\vb{p }, -\sigma } ) 
}{
    1
    + \frac{ u (\vb{p}) }{ 
        \hbar ( p_1^0 - p_4^0 ) 
        - 
        \left( 
            E_{\vb{p }\sigma } - E_{\vb{p }, -\sigma } 
        \right)
         - i\eta 
    } 
} 
. 
\end{split}
\end{eqnarray}
All other diagrams (which inevitably mix particle-particle and particle-hole ladders) vanish, an example of which is shown in Fig. \ref{fig:mix}. This is due to the restrictions on the occupations in Eq. (\ref{eq:pp}) and Eq. (\ref{eq:ph}), where we only have the combinations of $n_{\vb{p }\sigma } n_{\vb{p }, -\sigma }$, $(1-n_{\vb{p }\sigma }) (1-n_{\vb{p }, -\sigma } ) $ for the former,  and $(1-n_{\vb{p }\sigma })
    n_{\vb{p }, -\sigma } $, $n_{\vb{p }\sigma }
    (1-n_{\vb{p }, -\sigma } ) $ for the latter. 

\end{widetext}

\section{The Self-Energy Diagrams}
\label{Sec: Self-Energy}

In this section we study the electron self-energy $\Sigma_\sigma (\omega, \vb{k})$, especially its imaginary part, which tells us the decay rate and the broadening of the electron spectral function measured in the angle-resolved photoemission spectroscopy (ARPES):  
\begin{align} \label{eq:}
    \frac{1}{\tau} = \text{Im} \Sigma_\sigma (\omega, \vb{k}).
\end{align}
We evaluate the self-energy diagrams to the 2nd order of Hubbard interaction ($V^2$), which is the lowest order that gives rise to an imaginary part. We will, however, include {\it all} contributions from HKY interaction, using the ladder sums performed in the previous section.

The simplest diagram is the one with Hubbard interaction only (Fig. \ref{fig:_se_a}). Its imaginary part is 
\begin{widetext}
\begin{equation}	
\begin{split}
\text{Im }\Sigma_{\sigma}^{\text{\ref{fig:_se_a}}} (\omega, \vb{k})
=&
    V^2 
    \int \frac{d^2 k'}{(2\pi)^2} \frac{d^2 q}{(2\pi)^2} 
    \delta(
         \omega
        +
            E_{\vb{k'}, -\sigma } 
            + E_{\vb{k}'-\vb{q}, -\sigma } 
            - E_{ \vb{k}+ \vb{q}, \sigma }
    )
    \\& \qquad \times 
    \left[
        ( 1 - n_{ \vb{k}\sigma } )
        n_{ \vb{k} + \vb{q}, \sigma }
        ( 1 - n_{ \vb{k}', -\sigma } )
        n_{ \vb{k}'-\vb{q}, -\sigma }
     -
         n_{ \vb{k}\sigma }
        ( 1 - n_{ \vb{k} + \vb{q}, \sigma } )
        n_{ \vb{k}', -\sigma }
        ( 1 - n_{ \vb{k}'-\vb{q}, -\sigma } ) 
    \right]
    ,  
\end{split}
\label{eq: Fermi liquid result}
\end{equation}
\end{widetext} 
which yields the familiar Fermi liquid result near the (pseudo) Fermi surface:
\begin{equation}
\text{Im} \Sigma^{\ref{fig:_se_a}}_\sigma (\omega, \vb{k}) \approx  
-D^3 V^2 (\hbar \omega)^2,
\end{equation}
where $D$ is the density of states at the non-interacting Fermi level. We note however this results in much more broadening in the pseudogap region as the quasiparticle energy $\hbar \omega \sim |u|$ is bounded below by the size of the pseudogap, compared to that near the Fermi arcs where $\hbar \omega$ can be arbitrary small. This is consistent with the cuprate phenomenology.

\begin{widetext}
We now turn to the diagrams that involve HKY interaction ($H_2$). Those diagrams are in Fig. \ref{fig:_se}. Here we present the calculation on the self-energy diagram given by Fig. \ref{fig:_se_b} as an example. A full calculation on each diagram is presented in Appendix \ref{sec:app}.

\begin{figure}[!h]
    \centering
\begin{subfigure}[t]{0.45\textwidth} 
    \centering
    \includegraphics[width=3cm]{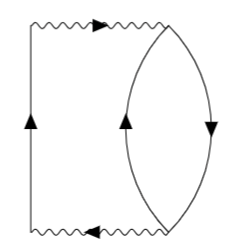}
    \caption{}
    \label{fig:_se_a}
\end{subfigure}
\\
\begin{subfigure}[t]{0.2\textwidth} 
    \centering
    \includegraphics[width=3cm]{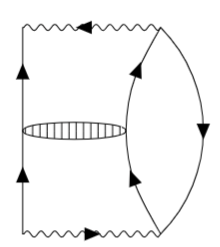}
    \caption{}
    \label{fig:_se_b}
\end{subfigure}
\begin{subfigure}[t]{0.2\textwidth} 
    \centering
    \includegraphics[width=3cm]{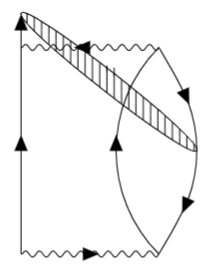}
    \caption{}
    \label{fig:_se_c}
\end{subfigure}
\begin{subfigure}[t]{0.2\textwidth} 
    \centering
    \includegraphics[width=3cm]{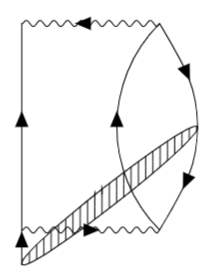}
    \caption{}
    \label{fig:_se_d}
\end{subfigure}
\begin{subfigure}[t]{0.2\textwidth} 
    \centering
    \includegraphics[width=3cm]{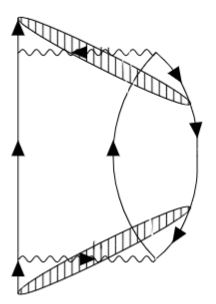}
    \caption{}
    \label{fig:_se_e}
\end{subfigure}
\\
\begin{subfigure}[t]{0.2\textwidth} 
    \centering
    \includegraphics[width=3cm]{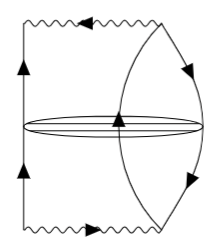}
    \caption{}
    \label{fig:_se_f}
\end{subfigure}
\begin{subfigure}[t]{0.2\textwidth} 
    \centering
    \includegraphics[width=3cm]{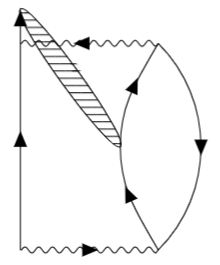}
    \caption{}
    \label{fig:_se_g}
\end{subfigure}
\begin{subfigure}[t]{0.2\textwidth} 
    \centering
    \includegraphics[width=3cm]{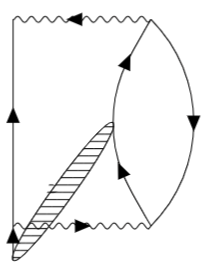}
    \caption{}
    \label{fig:_se_h}
\end{subfigure}
\begin{subfigure}[t]{0.2\textwidth} 
    \centering
    \includegraphics[width=3cm]{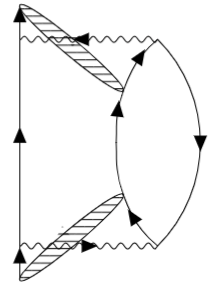}
    \caption{}
    \label{fig:_se_i}
\end{subfigure}
    \caption{Self-energy Feynman diagrams to the second order in Hubbard interaction ($V^2$). (a) is the self-energy diagrams in the second order with Hubbard interaction only. (b)-(i) are the self-energy diagrams with both Hubbard and the HKY ($H_2$) interactions.}
    \label{fig:_se}
\end{figure}

After performing the frequency integrals, the corresponding imaginary part of Fig. \ref{fig:_se_b} takes the form
\begin{equation} \label{eq:3b}
\begin{split}
\text{Im } \Sigma_{\sigma}^{\text{\ref{fig:_se_b}}} (\omega, \vb{k}) 
=&
i^6 \pi V^2 
\int \frac{d^2 q}{(2\pi)^2 }  \: 
\\& \times 
\left\{
    \left[
\delta(
    \omega + E_{-\vb{k}+2\vb{q}, -\sigma }
                - E_{\vb{q} \sigma }
                - E_{\vb{q}, -\sigma}
)
-
\delta(
    \omega + E_{-\vb{k}+2\vb{q}, -\sigma }
            - E_{\vb{q} \sigma }
            - E_{\vb{q}, -\sigma}
            - u_{\vb{q} }
\right]
(1 - n_{\vb{q} \sigma}) 
    (1 - n_{\vb{q}, -\sigma}) 
    n_{-\vb{k}+2\vb{q}, -\sigma }
\right. \\&\left. - 
\left[
\delta(
    \omega + E_{-\vb{k}+2\vb{q}, -\sigma }
                - E_{\vb{q} \sigma }
                - E_{\vb{q}, -\sigma}
)
-
\delta(
    \omega + E_{-\vb{k}+2\vb{q}, -\sigma }
            - E_{\vb{q} \sigma }
            - E_{\vb{q}, -\sigma}
            + u_{\vb{q} }
)
\right]
n_{\vb{q} \sigma}
    n_{\vb{q}, -\sigma} 
    ( 1 - n_{-\vb{k}+2\vb{q}, -\sigma } )
\right\}
,  
\end{split}
\end{equation}
Note compared to Eq. (\ref{eq: Fermi liquid result}), we have one {\em fewer} momentum integral to perform, despite the extra loop. This is due to the fact HKY interaction forces the propagators coupled by it to have the same momentum. This simplification changes the phase space constraints significantly and enhances $\text{Im } \Sigma$, as we demonstrate below.

To bring Eq. (\ref{eq:3b}) to a form closer to Eq. (\ref{eq: Fermi liquid result}), letting $-\vb{k}+2\vb{q}\rightarrow \vb{q}'$, we treat $\vb{q}'$ as an additional integration variable, compensated by an additional delta function. Rewrite the integrals in terms of the polar coordinates such that $(q_x, q_y) = (q\cos \phi, q\sin \phi)$ and $(q'_x, q_y') = (q'\cos \phi', q'\sin \phi')$. Eq. (\ref{eq:3b}) becomes
\begin{equation} \label{eq:3b_polar}
\begin{split}
\text{Im } \Sigma_{\sigma}^{\text{\ref{fig:_se_b}}} (\omega, \vb{k}) 
=&
i^6 \pi V^2 \frac{1}{(2\pi)^2}
\int qdq q'dq' d\phi d\phi' \: 
\frac{1}{q'}
\delta(q' - |-\vb{k}+2\vb{q}|)
\delta(\phi' - \phi_{-\vb{k}+2\vb{q}})
\\& \times 
\left\{
\left[
\delta(
    \omega + E_{\vb{q}', -\sigma }
                - 2\epsilon_{\vb{q}}
)
-
\delta(
    \omega + E_{\vb{q}', -\sigma }
            - 2\epsilon_{\vb{q}}
            - u_{\vb{q} }
)
\right]
\theta(\epsilon_{\vb{q}})
\theta(-E_{\vb{q}', -\sigma })
\right. \\& \left. - 
\left[
\delta(
    \omega + E_{\vb{q}', -\sigma }
                - 2\epsilon_{\vb{q}}
                - 2u_{\vb{q}}
)
-
\delta(
    \omega + E_{\vb{q}', -\sigma }
            - 2\epsilon_{\vb{q}}
            - 2u_{\vb{q}}
            + u_{\vb{q} }
)
\right]
\theta(-\epsilon_{\vb{q}}-u_{\vb{q}})
\theta(E_{\vb{q}', -\sigma })
\right\}
. 
\end{split}
\end{equation} 
We transform the integral from $\int qdq q'dq'$ to $\int d\epsilon dE'J(\epsilon, E', \phi, \phi')$ and $\int dE dE' d\phi d\phi'$, where $\epsilon=\epsilon_{\vb{q}}$, $E=\epsilon_{\vb{q}}+u_{\vb{q}}$, and $E' = E_{\vb{q}',-\sigma} $, and the Jacobian
\begin{align}
    & 
    J(\epsilon, E', \phi, \phi') = 
    q\pdv{q}{\epsilon} q'\pdv{q'}{E'} = D(\epsilon, \phi) D(E', \phi')
    ,\\&
    J(E, E', \phi, \phi') = 
    q\pdv{q}{E} q'\pdv{q'}{E'} = D(E, \phi) D(E', \phi')
   , 
\end{align}
with the angle-dependent density of states 
\begin{align}
& 
D(\epsilon, \phi) =  q \pdv{q}{\epsilon} 
,\\&
D(E, \phi) =  q \pdv{q}{E}
,\\&
D(E', \phi') =  q' \pdv{q'}{E'}
. 
\end{align}
In a two-dimensional system it is a good approximation to treat them as a constant in terms of density of states $D$: 
\begin{equation}
\begin{split}
    D(\epsilon, \phi) D(E', \phi')
    \approx 
    D(E, \phi) D(E', \phi')
    \approx 
    \left( 2\pi {D} \right)^2.
\end{split}
\end{equation}
Eq. (\ref{eq:3b_polar}) then becomes
\begin{equation} \label{eq:3b_E_phi}
\begin{split}
\text{Im }
&
\Sigma_{\sigma}^{\text{\ref{fig:_se_b}}} (\omega, \vb{k}) 
\\ \approx &
i^6 \pi D^2 
\left\{
\int d\epsilon dE' d\phi d\phi' 
\frac{1}{q'^2}
\delta\left(1 - \frac{|-\vb{k}+2\vb{q}|}{q'}\right)
\delta(\phi' - \phi_{-\vb{k}+2\vb{q}})
\left[
\delta(
    \omega + E'
                - 2\epsilon
)
-
\delta(
    \omega + E'
            - 2\epsilon
            + u_{\vb{q} }
)
\right]
\theta(\epsilon )
\theta(-E' )
\right. \\& \left. 
-
\int dE dE' d\phi d\phi' 
\frac{1}{q'^2}
\delta\left(1 - \frac{|-\vb{k}+2\vb{q}|}{q'}
\right)
\delta(\phi' - \phi_{-\vb{k}+2\vb{q}})
\left[
\delta(
    \omega + E'
                - 2E
)
-
\delta(
    \omega + E'
            - 2E
            + u_{\vb{q} }
)
\right]
\theta(-E)
\theta(E')
\right\}
, 
\end{split}
\end{equation}
where we have four integrals and three delta functions, and we let $\delta(q'-|-\vb{k}+2\vb{q}| ) \rightarrow
\frac{1}{q'}\delta\left(
    1 - \frac{|-\vb{k}+2\vb{q}|}{q'}
\right)$. We then consider the angle integral on $\phi'$. In order to express $|-\vb{k}+2\vb{q}|$, $q$, $q'$ in terms of the integral variables and their corresponding angle dependence, we solve the equation $E(k, \phi_{\vb{k}})=E$ so that $k=f(E, \phi)$. Rewriting $u_{\vb{q}}=u_{q\phi}$, the approximate Eq. (\ref{eq:3b_E_phi}) becomes 
\begin{equation} \label{eq:3b_f}
\begin{split}
\text{Im }
\Sigma_{\sigma}^{\text{\ref{fig:_se_b}}} (\omega, \vb{k}) 
\approx &
i^6 \pi D^2 
\int d\phi' 
    \delta(\phi' - \phi_{-\vb{k}+2\vb{q}} )
\\& \times 
\left\{
\int d\epsilon dE' d\phi
\frac{1}{f^2(E', \phi')}
    \delta\left[
    1 - \frac{
        f(\epsilon, \phi_{-\vb{k}+2\vb{q}})
    }{
        f(E', \phi')
        }
    \right]
\left[
\delta(
    \omega + E'
                - 2\epsilon
)
-
\delta(
    \omega + E'
            - 2\epsilon
            + u_{f(\epsilon, \phi), \phi}
)
\right]
\theta(\epsilon )
\theta(-E' )
\right. \\& \left. 
-
\int dE dE' d\phi
\frac{1}{f^2(E', \phi')}
    \delta\left[
    1 - \frac{
        f(E, \phi_{-\vb{k}+2\vb{q}})
    }{
        f(E', \phi')
        }
    \right]
\left[
\delta(
    \omega + E'
                - 2E
)
-
\delta(
    \omega + E'
            - 2E
            + u_{f(E, \phi), \phi}
)
\right]
\theta(-E)
\theta(E')
\right\}
. 
\end{split}
\end{equation}
Performing the angle integral over $\phi'$ yields 
\begin{equation} \label{eq:3b_f_integrate_phi'}
\begin{split}
\text{Im }
&
\Sigma_{\sigma}^{\text{\ref{fig:_se_b}}} (\omega, \vb{k}) 
\approx \\&
i^6 \pi D^2 
\left\{
\int d\epsilon dE' d\phi 
 \frac{1}{f^2(E', \phi_{-\vb{k}+2\vb{q}})}
    \delta\left[
    1 - \frac{
        f(\epsilon, \phi_{-\vb{k}+2\vb{q}})
    }{
        f(E', \phi_{-\vb{k}+2\vb{q}})
        }
    \right]
\left[
\delta(
    \omega + E'
                - 2\epsilon
)
-
\delta(
    \omega + E'
            - 2\epsilon
            + u_{f(\epsilon, \phi), \phi}
)
\right]
\theta(\epsilon )
\theta(-E' )
\right. \\& \left. 
-
\int dE dE' d\phi 
\frac{1}{f^2(E', \phi_{-\vb{k}+2\vb{q}})}
    \delta\left[
    1 - \frac{
        f(E, \phi_{-\vb{k}+2\vb{q}})
    }{
        f(E', \phi_{-\vb{k}+2\vb{q}})
        }
    \right]
\left[
\delta(
    \omega + E'
                - 2E
)
-
\delta(
    \omega + E'
            - 2E
            + u_{f(E, \phi), \phi}
)
\right]
\theta(-E)
\theta(E')
\right\}
. 
\end{split}
\end{equation}
We notice that $\phi_{-\vb{k}+2\vb{q}}$ can by further replaced by another function $g$ in terms of $\vb{k}$, $\epsilon$, $E$ and $\phi$. (\ref{eq:3b_f_integrate_phi'}) becomes 
\begin{equation} \label{eq:3b_f_phi}
\begin{split}
\text{Im }
&
\Sigma_{\sigma}^{\text{\ref{fig:_se_b}}} (\omega, \vb{k}) 
\approx \\ &
i^6 \pi D^2 
\left\{
\int d\epsilon dE' d\phi 
    \frac{1}{f^2[E', g(\vb{k}, \epsilon, \phi)]}
    \delta\left\{
    1 - \frac{
        f[\epsilon, g(\vb{k}, \epsilon, \phi)]
    }{
        f[E', g(\vb{k}, \epsilon, \phi)]
        }
    \right\}
\left[
\delta(
    \omega + E'
                - 2\epsilon
)
-
\delta(
    \omega + E'
            - 2\epsilon
            + u_{f(\epsilon, \phi), \phi}
)
\right]
\theta(\epsilon )
\theta(-E' )
\right. \\& \left. 
-
\int dE dE' d\phi 
    \frac{1}{f^2[E', g(\vb{k}, E, \phi)]}
    \delta\left\{
    1 - \frac{
        f[E, g(\vb{k}, E, \phi)]
    }{
        f[E', g(\vb{k}, E, \phi)]
        }
    \right\}
\left[
\delta(
    \omega + E'
                - 2E
)
-
\delta(
    \omega + E'
            - 2E
            + u_{f(E, \phi), \phi}
)
\right]
\theta(-E)
\theta(E')
\right\}
. 
\end{split}
\end{equation}
We first perform the integral over $\phi$. The remaining delta functions give us $\phi$'s dependence on $\vb{k}$, $\epsilon$, $E$ and $E'$, and we denote it as a function $h$. Hence, (\ref{eq:3b_f_phi}) becomes 
\begin{equation} \label{eq:3b_integrate_phi}
\begin{split}
\text{Im }
&
\Sigma_{\sigma}^{\text{\ref{fig:_se_b}}} (\omega, \vb{k}) 
\approx \\ &
i^6 \pi D^2 
\left\{
\int d\epsilon dE'
    \frac{1}{
        f^2\{E', g[
                \vb{k}, \epsilon, h(\vb{k}, \epsilon, E')]\}
    }
\left[
\delta(
    \omega + E'
                - 2\epsilon
)
-
\delta(
    \omega + E'
            - 2\epsilon
            + u_{f[\epsilon, h(\vb{k}, \epsilon, E')], h(\vb{k}, \epsilon, E')} 
)
\right]
\theta(\epsilon )
\theta(-E' )
\right. \\& \left. 
-
\int dE dE'
    \frac{1}{
        f^2\{E', g[
                \vb{k}, E, h(\vb{k}, E, E')]
        \}
    }
\left[
\delta(
    \omega + E'
                - 2E
)
-
\delta(
    \omega + E'
            - 2E
            + u_{f[E, h(\vb{k}, E, E')], h(\vb{k}, E, E')}
)
\right]
\theta(-E)
\theta(E')
\right\}
. 
\end{split}
\end{equation}
While we do not know the exact form of the terms $1/f^2$, they give us quantities of order $1/(\text{Fermi momentum})^2$, which is of order $O(1)$ for generic lattice filling. Its dependence on $\epsilon$, $E$, $E'$ and $\vb{k}$ is unimportant due to the phase space constraints, as will become clear soon. We are then left with integrals with energy variables $\epsilon$, $E$ and $E'$:
\begin{equation} \label{eq:3b_E}
\begin{split}
\text{Im } \Sigma_{\sigma}^{\text{\ref{fig:_se_b}}} (\omega, \vb{k}) 
\approx &
i^6 \pi V^2 D^2 
\left\{
\int d\epsilon dE' 
\left[
\delta(
    \omega + E' 
            - 2 \epsilon
)
-
\delta(
    \omega + E'
            - 2\epsilon
            - u_{f[\epsilon, h(\vb{k}, \epsilon, E')], h(\vb{k}, \epsilon, E')} 
)
\right]
\theta(\epsilon)
\theta(-E' )
\right. \\& \left. - 
\int dE dE' 
\left[
\delta(
    \omega + E' - 2E
)
-
\delta(
    \omega + E'
            - 2E
            + u_{f[E, h(\vb{k}, E, E')], h(\vb{k}, E, E')} 
)
\right]
\theta(-E )
\theta(E' )
\right\}
, 
\end{split}
\end{equation} 
We can then carry out the integral by assuming $u_{f[\epsilon, h(\vb{k}, \epsilon, E')], h(\vb{k}, \epsilon, E')} \approx u_{f[E, h(\vb{k}, E, E')], h(\vb{k}, E, E')}  \approx |u|$, where $|u|$ is some constant average over $u_{f[\epsilon, h(\vb{k}, \epsilon, E')], h(\vb{k}, \epsilon, E')} $ or $u_{f[E, h(\vb{k}, E, E')], h(\vb{k}, E, E')} $. The integrals over $E'$ give 
\begin{equation}
\begin{split}
\text{Im }\Sigma_{\sigma}^{\text{\ref{fig:_se_b}}} (\omega, \vb{k})
\approx 
i^6 \pi V^2 D^2 
\left\{
\int d\epsilon \theta(\epsilon)
    \left[
    \theta(-2\epsilon+\omega)
    - 
    \theta(-2\epsilon+\omega-|u|)
    \right]
    - 
\int dE \theta(E)
    \left[
    \theta(-2E+\omega)
    - 
    \theta(-2E+\omega-|u|)
    \right]
\right\}
.
\end{split}
\end{equation}
The integrals on $\epsilon$, $E$ give 
\begin{align} \label{eq:3b_final}
\text{Im } \Sigma_{\sigma}^{\text{\ref{fig:_se_b}}} (\omega, \vb{k}) 
\approx &  
i^6 \pi V^2
\left[
    \left(
    \frac{\omega}{2} - 
    \frac{\omega - |u|}{2}
    \right)
    - 
    \left(
    \frac{\omega}{2} - 
    \frac{\omega + |u|}{2}
    \right)
\right]
\\=& 
i^6 \pi V^2 D^2 
\left[
    \frac{|u|}{2}
    - 
    \left(
        -\frac{|u|}{2}
    \right)
\right]
\\=&  
i^6 \pi V^2 D^2 |u|
. 
\end{align}
Here we note that the linearity in $|u|$ comes from the energy integral per the phase space restrictions given by the step functions $\theta(-2\epsilon+\omega)$, $\theta(-2\epsilon+\omega-u_{\vb{q}})$, $\theta(E)$, and $\theta(-2E+\omega+u_{\vb{q}})$. The final result would not be exactly linear in $|u|$. Hence, we append a function $f(\vb{k})$ varying in $\vb{k}$: 
\begin{equation} \label{eq:3b_final_k}
\begin{split}
\text{Im } \Sigma_{\sigma}^{\text{\ref{fig:_se_b}}} (\omega, \vb{k}) 
\approx &  
-\pi V^2 D^2 |u| f(\vb{k})
, 
\end{split}
\end{equation}
where $f(\vb{k})$ is a dimensionless quantity of order $O(1)$. Letting $\omega \rightarrow E_{\vb{k}\sigma}$, we obtain the self-energy results from other diagrams: 
\begin{align}
& 
\text{Im } \Sigma_{\sigma}^{\text{\ref{fig:_se_f}}} (\vb{k}) 
\approx 
- \frac{\pi}{2} V^2 D^2 u_{\vb{k}} (n_{\vb{k}\sigma}-n_{\vb{k}, -\sigma } )
\label{eq:se_f_final}
,\\& 
\text{Im }\Sigma_{\sigma}^{\text{\ref{fig:_se_g}}} (\vb{k})
\approx 
\text{Im}\Sigma_{\sigma}^{\text{\ref{fig:_se_h}}} (\vb{k})
\approx 
\text{Im }\Sigma_{\sigma}^{\text{\ref{fig:_se_i}}} (\vb{k})
\approx 
- \pi V^2 D^2 u_{\vb{k}}
(n_{\vb{k}\sigma} - n_{\vb{k}, -\sigma } )^2  
\label{eq:se_hi_final}
. 
\end{align}
The rest imaginary parts resulting from Fig. \ref{fig:_se_c}, \ref{fig:_se_d}, and \ref{fig:_se_e} share the same form but appear more complicated as we presented later in the expressions (\ref{eq:app_c_omegaE}) and (\ref{eq:app_e_omegaE}). However, since we are only interested in the cases that occur near Fermi-arcs and pseudo-Fermi surfaces, further simplifications can be done so that overall these imaginary parts results in zero or a quantity linear in $u_{\vb{k}}$.

\end{widetext}

\section{Summary and Discussions}
\label{sec: Summary}

In this paper we studied the model introduced in Ref. \cite{Yang21} (referred to as HKY model) which gives rise to Fermi arcs and pseudogap, perturbed by Hubbard interaction. We found the combination of Hubbard and HKY interactions gives rise to a non-zero imaginary part to the electron self-energy in the low-energy limit. The origin of such non-Fermi liquid behavior lies in the singular nature of HKY interaction, which has infinite range in real space.

While our work was motivated by the cuprates, and gives rise to results that are qualitatively consistent with its phenomenology, the specific (and certainly over-simplified) model we studied should not be taken as a realistic description of the physics of cuprates. Its value lies, instead, in its simplicity, which demonstrates not only the possibility of Fermi arcs and pseudogap, but also that they go hand-in-hand with each other and with the observed non-Fermi liquid behavior. In fact recent years have witnessed increasing activities in research using models that are extensions of the HK model \cite{phillips20, li, nesselrodt, zhong} aimed at understanding cuprate phenomenology, including superconductivity itself. It is our hope that our work provides a starting point to build more realistic models for cuprates and other strongly correlated electron systems.

\section*{Acknowledgments}
This work was supported by the National Science Foundation Grant No. DMR-1932796, and performed at the National High Magnetic Field Laboratory, which is supported by National Science Foundation Cooperative Agreement No. DMR-1644779, and the State of Florida.


\begin{widetext}
\appendix
\section{Calculations on the self-energy diagrams}
Here in the appendix we present more calculation details for each self energy diagram in the section \label{sec:app}.

\begin{figure}[!h]
    \centering
\begin{subfigure}[t]{0.45\textwidth} 
    \centering
    \includegraphics[width=8cm]{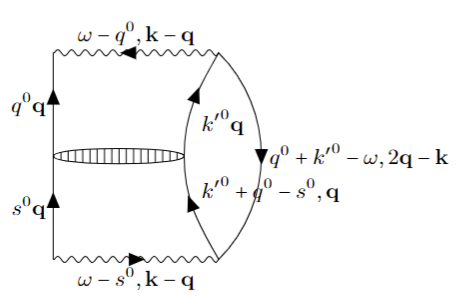}
    \caption*{Fig. \ref{fig:_se_b}}
\end{subfigure}
\begin{subfigure}[t]{0.45\textwidth} 
    \centering
    \includegraphics[width=7cm]{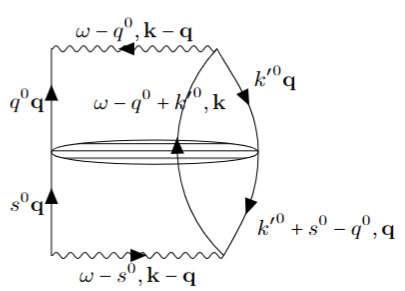}
    \caption*{Fig. \ref{fig:_se_f}}
\end{subfigure}
    \label{fig:se1a}
\end{figure}

The self-energy terms given by Fig. \ref{fig:_se_b} and Fig. \ref{fig:_se_b} are 
\begin{equation}	
\begin{split}
\Sigma_{\sigma}^{\text{\ref{fig:_se_b}}} (\omega, \vb{k})
=&
i^3 V^2
\int \frac{d^2 q }{(2\pi )^2 } 
\frac{ds^0 }{(2\pi )} \frac{dq^0 }{(2\pi )} \frac{d{k'}^0 }{(2\pi) } 
G_\sigma^0 (q^0 \vb{q} ) G_\sigma^0 ( s^0 \vb{q} )
G_{-\sigma}^0 ( k'^0 \vb{q})
\\ & \times 
    G_{-\sigma}^0 ( k'^0+q^0-s^0, \vb{q}) 
    G_{-\sigma}^0 ( q^0+k'^0-\omega, 2\vb{q}-\vb{k})
\Gamma_{\text{pp}} (q^0+k'^0, \vb{q} )
. 
\end{split}
\end{equation}
The frequency integral gives 
\begin{equation}
\begin{split}
\Sigma_{\sigma}^{\text{\ref{fig:_se_b}}} (\omega, \vb{k})
=&
i^6 V^2 
\int \frac{d^2 q}{(2\pi)^2 } 
\\& \times 
\left[
n_{\vb{q} \sigma}
    n_{\vb{q}, -\sigma} 
    ( 1 - n_{-\vb{k}+2\vb{q}, -\sigma } )
\left(
\frac{1}{
    \omega + E_{-\vb{k}+2\vb{q}, -\sigma }
            - E_{\vb{q} \sigma }
            - E_{\vb{q}, -\sigma}
            - i \eta 
}
- 
\frac{1}{
    \omega + E_{-\vb{k}+2\vb{q}, -\sigma }
            - E_{\vb{q} \sigma }
            - E_{\vb{q}, -\sigma}
            + u_{\vb{q} }
            - i \eta 
}
\right)
\right.  \\& \left. +
(1 - n_{\vb{q} \sigma}) 
    (1 - n_{\vb{q}, -\sigma}) 
    n_{-\vb{k}+2\vb{q}, -\sigma }
\left(
\frac{1}{
    \omega + E_{-\vb{k}+2\vb{q}, -\sigma }
        - E_{\vb{q} \sigma }
        - E_{\vb{q}, -\sigma}
            + i \eta
}
- 
\frac{1}{
     \omega + E_{-\vb{k}+2\vb{q}, -\sigma }
            - E_{\vb{q} \sigma }
            - E_{\vb{q}, -\sigma}
            - u_{\vb{q} }
            + i \eta 
}
\right)
\right] 
\end{split}
\end{equation}
The corresponding imaginary part and a sample calculation on Fig. \ref{fig:_se_b} is shown in Section \ref{Sec: Self-Energy}. The final result is 
\begin{equation} \label{eq:app_3b_final_k}
\begin{split}
\text{Im } \Sigma_{\sigma}^{\text{\ref{fig:_se_b}}} (\omega, \vb{k}) 
\approx &  
-\pi V^2 D^2 |u| f(\vb{k})
, 
\end{split}
\end{equation}
where $f(\vb{k})$ is a dimensionless quantity of order $O(1)$.


Next we would like to present the calculation on Fig. \ref{fig:_se_f}. 
\begin{equation}	
\begin{split}
\Sigma_{\sigma}^{\text{\ref{fig:_se_f}}} (\omega, \vb{k})
=&
i^3 V^2
\int \frac{d^2 q }{(2\pi )^2 } 
\frac{ds^0 }{(2\pi )} \frac{dq^0 }{(2\pi )} \frac{d{k'}^0 }{(2\pi) } 
G_\sigma^0 (q^0 \vb{q} ) G_\sigma^0 ( s^0 \vb{q} )
G_{-\sigma}^0 ( k'^0 \vb{q})
 \\ & \times 
    G_{-\sigma}^0 ( k'^0-q^0+s^0, \vb{q}) 
    G_{-\sigma}^0 ( -q^0+k'^0+\omega, \vb{k})
\Gamma_{\text{ph}} (q^0-   k'^0, \vb{q} )
. 
\end{split}
\end{equation}
The frequency integral gives 
\begin{equation}
\begin{split}
\Sigma_{\sigma}^{\text{\ref{fig:_se_f}}} (\omega, \vb{k})
=&
i^6 V^2 
\int \frac{d^2 q}{(2\pi)^2 } 
\\& \times 
\left[
(1- n_{\vb{q} \sigma})
    n_{\vb{q}, -\sigma} 
    ( 1 - n_{\vb{k}, -\sigma } )
\left(
\frac{1}{
    \omega 
        - E_{\vb{k}, -\sigma }
        - E_{\vb{q} \sigma }
        + E_{\vb{q}, -\sigma}
        + i \eta
}
- 
\frac{1}{
    \omega 
        - E_{\vb{k}, -\sigma }
        - E_{\vb{q} \sigma }
        + E_{\vb{q}, -\sigma}
        -               u_{\vb{q} }
        + i \eta 
}
\right)
\right.  \\& \left. +
n_{\vb{q} \sigma}
    (1 - n_{\vb{q}, -\sigma}) 
    n_{\vb{k}, -\sigma }
\left(
\frac{1}{
    \omega 
        - E_{-\vb{k}, -\sigma }
        - E_{\vb{q} \sigma }
        + E_{\vb{q}, -\sigma}
        - i \eta 
}
- 
\frac{1}{
    \omega 
        + E_{\vb{k}, -\sigma }
        - E_{\vb{q} \sigma }
        + E_{\vb{q}, -\sigma}
        + u_{\vb{q} }
        - i \eta 
}
\right)
\right]
\end{split}
\end{equation}
The corresponding imaginary part is 
\begin{equation} \label{eq:se1bim2}
\begin{split}
\text{Im } \Sigma_{\sigma}^{\text{\ref{fig:_se_f}}} (\omega, \vb{k}) 
=&
i^6 \pi V^2 
\int \frac{d^2 q}{(2\pi)^2 }  \: 
\left\{
\left[
- 
\delta(
    \omega 
        - E_{\vb{k}, -\sigma }
        - E_{\vb{q}\sigma }
        + E_{\vb{q}, -\sigma }
)
+
\delta(
    \omega 
        - E_{\vb{k}, -\sigma }
        - E_{\vb{q}\sigma }
    + E_{\vb{q}, -\sigma }
        - u_{\vb{q} }
)
\right]
(1- n_{\vb{q} \sigma})
    n_{\vb{q}, -\sigma} 
    ( 1 - n_{\vb{k}, -\sigma } )
\right. \\& \left. +
\left[
\delta(
\omega 
        - E_{\vb{k}, -\sigma }
        - E_{\vb{q}\sigma }
    + E_{\vb{q}, -\sigma }
)
-
\delta(
    \omega 
        - E_{\vb{k}, -\sigma }
        - E_{\vb{q}\sigma }
    + E_{\vb{q}, -\sigma }
        + u_{\vb{q} }
)
\right]
n_{\vb{q} \sigma}
    (1 - n_{\vb{q}, -\sigma}) 
    n_{\vb{k}, -\sigma }
\right\}
. 
\end{split}
\end{equation}
After simplification we have 
\begin{equation}
\begin{split} \label{eq:se_f_mom}
\text{Im } \Sigma_{\sigma}^{\text{\ref{fig:_se_f}}} (\omega, \vb{k}) 
=&
i^3 \pi V^2 
\int \frac{d^2 q}{(2\pi)^2 }  \: 
\left\{
\left[
-
\delta(
    \omega 
        - E_{\vb{k}, -\sigma }
       - u_{\vb{q}}
)
+
\delta(
    \omega 
        - E_{\vb{k}, -\sigma }
        - 2 u_{\vb{q}}
)
\right]
\theta(\epsilon_{\vb{q}}+ u_{\vb{q}})
\theta(-\epsilon_{\vb{q}})
(1-n_{\vb{k},-\sigma})
\right. \\& \left. -
\left[
\delta(
\omega 
        - E_{\vb{k}, -\sigma }
        + u_{\vb{q}}
)
-
\delta(
    \omega 
        - E_{\vb{k}, -\sigma }
        + 2u_{\vb{q}}
)
\right]
\theta(-\epsilon_{\vb{q}})
\theta(\epsilon_{\vb{q}}+u_{\vb{q}})
n_{\vb{k},-\sigma}
\right\}
. 
\end{split}
\end{equation}
We let $E_{\vb{q}'} = \epsilon_{\vb{q}}+u_{\vb{q}}$ such that $E_{q'\phi'}=\epsilon_{q\phi}+u_{q\phi}$ in polar coordinates. We solve this equation in favor of $q'$ and $\phi'$:
\begin{align}
    & 
    q' = f(\epsilon_{q\phi}+u_{q\phi}, \phi')
    , \\
    &
    \phi' = g(\epsilon_{q\phi}+u_{q\phi}, q') = g[\epsilon_{q\phi}+u_{q\phi}, f(\epsilon_{q\phi}+u_{q\phi}, \phi')]
    . 
\end{align}
Eq. (\ref{eq:se_f_mom}) becomes 
\begin{equation}
\begin{split} \label{eq:se_f_delta}
\text{Im } \Sigma_{\sigma}^{\text{\ref{fig:_se_f}}} (\omega, \vb{k}) 
=&
i^6 \frac{\pi}{(2\pi)^2} V^2 
\int qdq q'dq' d\phi d\phi' \:
\frac{1}{q'^2}
\delta\left[
    1 - \frac{f(\epsilon_{q\phi}+u_{q\phi}, \phi')}{q'}
\right]
\delta\left\{
    g[
        E_{q'\phi'}, f(\epsilon_{q\phi}+u_{q\phi}, \phi')
    ]
\right\}
\\& \times 
\left\{
\left[
-
\delta(
    \omega 
        - E_{\vb{k}, -\sigma }
       - u_{\vb{q}}
)
+
\delta(
    \omega 
        - E_{\vb{k}, -\sigma }
        - 2 u_{\vb{q}}
)
\right]
\theta(\epsilon_{\vb{q}}+ u_{\vb{q}})
\theta(-\epsilon_{\vb{q}})
(1-n_{\vb{k},-\sigma})
\right. \\& \left. +
\left[
\delta(
\omega 
        - E_{\vb{k}, -\sigma }
        + u_{\vb{q}}
)
-
\delta(
    \omega 
        - E_{\vb{k}, -\sigma }
        + 2u_{\vb{q}}
)
\right]
\theta(-\epsilon_{\vb{q}})
\theta(\epsilon_{\vb{q}}+u_{\vb{q}})
n_{\vb{k},-\sigma}
\right\}
,  
\end{split}
\end{equation}
where 
\begin{align}
\int d\phi d\phi' \:
\frac{1}{q'^2}
\delta\left[
    1 - \frac{f(\epsilon_{q\phi}+u_{q\phi}, \phi')}{q'}
\right]
\delta \left\{
    \phi' - 
    g[
        E_{q'\phi'}, f(\epsilon_{q\phi}+u_{q\phi}, \phi')
    ]
\right\}
\end{align}
gives a quantity of the order $1/(\text{Fermi momentum})^2$ and thus of the order $O(1)$. We then transform $\int qdq q'dq'$ to $\int D(E, \phi) D(E', \phi') dE dE'$ where $E=\epsilon_{\vb{q}}$ and $E' = \epsilon_{\vb{q}}+u_{\vb{q}}$ such that $u_{\vb{q}}=E-E'$, and we let $D(E, \phi) D(E', \phi') \approx D^2 $. Eq. (\ref{eq:se_f_delta}) becomes 
\begin{equation}
\begin{split} \label{eq:se_f_E}
\text{Im } \Sigma_{\sigma}^{\text{\ref{fig:_se_f}}} (\omega, \vb{k}) 
\approx 
i^6 \pi V^2 D^2  
\int dE dE' 
\: &
\left\{
\left[
-
\delta(
    \omega 
        - E_{\vb{k}, -\sigma }
       - E' + E
)
+
\delta(
    \omega 
        - E_{\vb{k}, -\sigma }
        - 2 E' - 2E 
)
\right]
\theta(E')
\theta(-E)
(1-n_{\vb{k},-\sigma})
\right. \\-& \left. 
\left[
\delta(
\omega 
        - E_{\vb{k}, -\sigma }
        + E'-E
)
-
\delta(
    \omega 
        - E_{\vb{k}, -\sigma }
        + 2E' - 2E
)
\right]
\theta(-E)
\theta(E')
n_{\vb{k},-\sigma}
\right\}
.  
\end{split}
\end{equation}
This gives 
\begin{align}
\text{Im } \Sigma_{\sigma}^{\text{\ref{fig:_se_f}}} (\omega, \vb{k}) 
\approx & 
i^6 \frac{\pi}{2} V^2 D^2 
(E_{\vb{k},-\sigma} - \omega )
. 
\end{align}
As $\omega \rightarrow E_{\vb{k}\sigma}$, we obtain 
\begin{align}
\text{Im } \Sigma_{\sigma}^{\text{\ref{fig:_se_f}}} (\vb{k}) 
\approx & 
-\frac{\pi}{2} V^2 D^2 u_{\vb{k}}
(n_{\vb{k}\sigma} - n_{\vb{k},-\sigma})
. 
\end{align}


We consider Fig. \ref{fig:_se_c}, \ref{fig:_se_g}, \ref{fig:_se_d}, and \ref{fig:_se_h} together. We find the equality in Fig. \ref{fig:_se_cdgh}.
\begin{figure}
\center
\includegraphics[width=10cm]{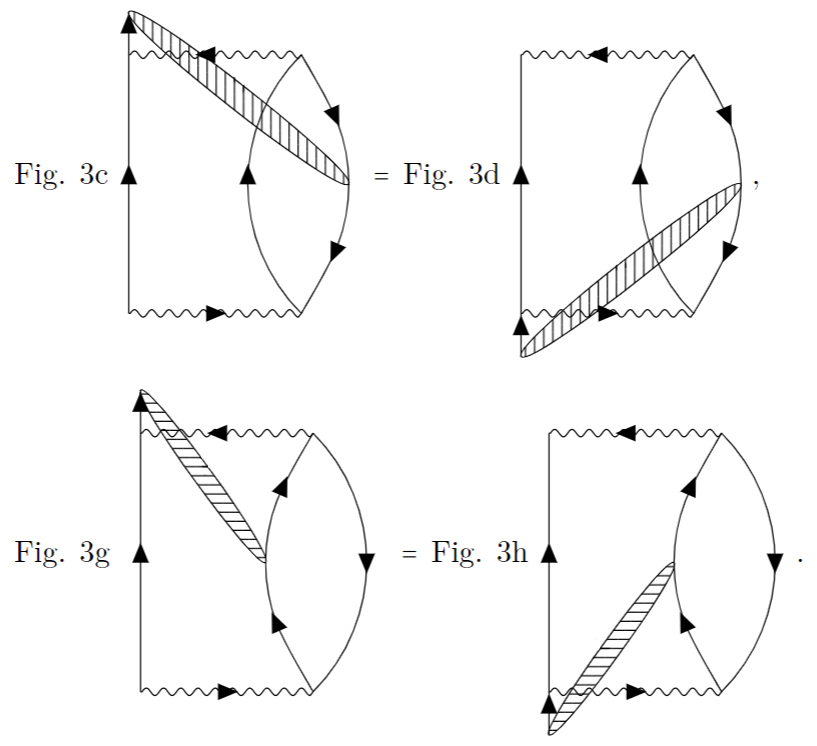}
\caption{}
\label{fig:_se_cdgh}
\end{figure}
So here we only consider Fig. \ref{fig:_se_c} and \ref{fig:_se_g}. We start with Fig. \ref{fig:_se_c} first 
\begin{equation}
\begin{split}
\Sigma_{\sigma}^{\text{\ref{fig:_se_c}}} (\omega, \vb{k})
=&
    i^3 V^2
    \int \frac{d^2 q}{(2\pi)^2 } 
    \int \frac{ds^0}{2\pi} \frac{dq^0}{2\pi} \frac{dk'^0}{2\pi} 
    G_\sigma^0 (s^0, \vb{k}) 
    G_\sigma^0 (s^0-q^0, \vb{k}-\vb{q})
    G_{-\sigma}^0 ({k'}^0, \vb{k}) 
    \\ & \times
    G_{-\sigma}^0 ({k'}^0-\omega+s^0, \vb{k})
    G_{-\sigma}^0 ({k'}^0+q^0, \vb{k}+\vb{q})
    \Gamma_{\text{pp}} (s^0 + {k'}^0, \vb{k})
    . 
\end{split}
\end{equation}
\begin{figure}[!h]
    \centering
\begin{subfigure}[t]{0.45\textwidth} 
    \centering
    \includegraphics[width=6cm]{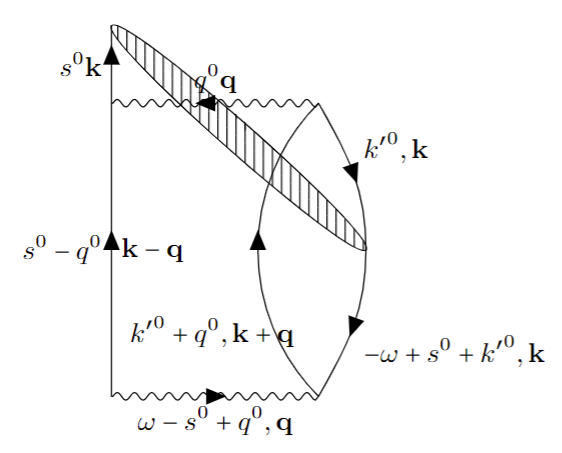}
    \caption*{Fig. \ref{fig:_se_c}}
    \label{fig:se_c}
\end{subfigure}
\begin{subfigure}[t]{0.45\textwidth} 
    \centering
    \includegraphics[width=6cm]{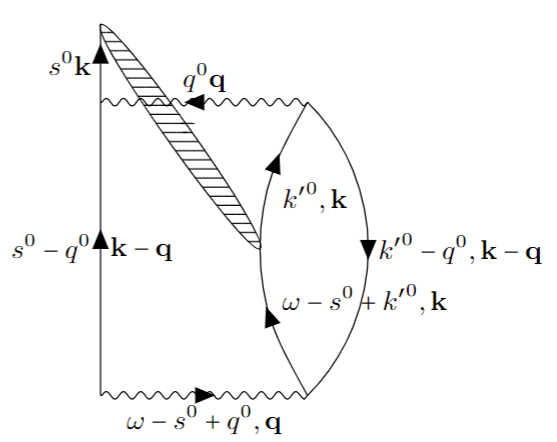}
    \caption*{Fig. \ref{fig:_se_g}}
    \label{fig:se_g}
\end{subfigure}
    \label{fig:se_cg}
\end{figure}
The frequency integral gives 
\begin{equation}
\begin{split}
\Sigma_{\sigma}^{\text{\ref{fig:_se_c}}} (\omega, \vb{k})
=&
i^6 V^2 u_{\vb{k} }
\int \frac{d^2 q}{(2\pi)^2}
\\& \times
\left[
 -
    \frac{
        (1 - n_{\vb{k}, -\sigma } ) 
        (1 - n_{\vb{k}\sigma } )
        n_{\vb{k}+ \vb{q}, -\sigma }
        n_{\vb{k}-\vb{q}, \sigma } 
    }{
        (
            - \omega 
            + E_{\vb{k}-\vb{q}, -\sigma } 
            + E_{\vb{k}+\vb{q}, -\sigma } 
            - E_{\vb{k}, -\sigma } 
            + i\eta 
        )
        (
            - E_{\vb{k}\sigma} 
            + E_{\vb{k}-\vb{q}, -\sigma } 
            + E_{\vb{k}+\vb{q}, -\sigma } 
            - E_{\vb{k}, -\sigma } 
            - u_{\vb{k}}
            + i\eta 
        )
    }
\right. \\& \left. \quad 
+
    \frac{
        n_{\vb{k}, -\sigma } 
        n_{\vb{k}\sigma }
        (1 - n_{\vb{k}+ \vb{q}, -\sigma })
        (1 - n_{\vb{k}-\vb{q}, \sigma } )
    }{
        (
            - \omega 
            + E_{\vb{k}-\vb{q}, -\sigma } 
            + E_{\vb{k}+\vb{q}, -\sigma } 
            - E_{\vb{k}, -\sigma } 
            - i\eta 
        )
        (
            - E_{\vb{k}\sigma} 
            + E_{\vb{k}-\vb{q}, -\sigma } 
            + E_{\vb{k}+\vb{q}, -\sigma } 
            - E_{\vb{k}, -\sigma }
            + u_{\vb{k}}
            - i\eta 
        )
    }
\right]
. 
\end{split}
\end{equation}
The imaginary part is 
\begin{equation} \label{eq:app_3c}
\begin{split}
\text{Im }\Sigma_{\sigma}^{\text{\ref{fig:_se_c}}} (\omega, \vb{k})
= &
-\pi V^2 
u_{\vb{k}}
\int \frac{d^2 q}{(2\pi)^2}
\delta 
    (
        \omega 
            - E_{\vb{k}-\vb{q}, \sigma } 
            - E_{\vb{k}+\vb{q}, -\sigma } 
            + E_{\vb{k}, -\sigma } 
    )
\\& \times 
\left[
\text{sgn} \left(
    -\omega - E_{\vb{k}\sigma} - 2E_{\vb{k}, -\sigma} 
    + 2E_{\vb{k}-\vb{q}, \sigma }
    + 2E_{\vb{k}+\vb{q}, -\sigma }
    - u_{\vb{k}}
\right)
\frac{
(1 - n_{\vb{k}, -\sigma } ) 
        (1 - n_{\vb{k}\sigma } )
        n_{\vb{k}+ \vb{q}, -\sigma }
        n_{\vb{q}, \sigma } 
}{
    |- E_{\vb{k}, -\sigma } 
    + E_{\vb{k}+\vb{q}, -\sigma }
    + E_{\vb{k}-\vb{q}, \sigma }
    - E_{\vb{k}\sigma}
    + u_{\vb{k}}|
}
\right. \\& \left. 
+ 
\text{sgn} \left(
    -\omega - E_{\vb{k}\sigma} - 2E_{\vb{k}, -\sigma} 
    + 2E_{\vb{k}-\vb{q}, \sigma }
    + 2E_{\vb{k}+\vb{q}, -\sigma }
    + u_{\vb{k}}
\right)
\frac{
    n_{\vb{k}, -\sigma } 
        n_{\vb{k}\sigma }
        (1 - n_{\vb{k}+ \vb{q}, -\sigma })
        (1 - n_{\vb{k}-\vb{q}, \sigma } )
}{
    |- E_{\vb{k}, -\sigma } 
    + E_{\vb{k}+\vb{q}, -\sigma }
    + E_{\vb{k}-\vb{q}, \sigma }
    - E_{\vb{k}\sigma}
    - u_{\vb{k}}|
}
\right]
.
\end{split}
\end{equation}
We let $\vb{k}+\vb{q}\rightarrow \vb{q}'$ and treat $\vb{q}'$ as an additional integration variable. We rewrite the integrals in terms of the polar coordinates such that $(q_x, q_y) = (q\cos \phi, q\sin \phi)$ and $(q'_x, q_y') = (q'\cos \phi', q'\sin \phi')$, and we transform the integral $\int q dq q' dq' \rightarrow \int (2\pi D)^2 dE dE' J(E, E')$, where $E = E_{\vb{k}-\vb{q}, \sigma}$ and  $E'=E_{\vb{q}', -\sigma}$. After we ignore the quantity of the order $O(1)$, the expression (\ref{eq:app_3c}) then becomes 
\begin{equation}
\begin{split} \label{eq:se_c_E_phi}
\text{Im}
\Sigma_{\sigma}^{\text{\ref{fig:_se_c}}} 
(\omega, \vb{k})
\approx
&
-\pi V^2 
u_{\vb{k}} D^2 
\int dE dE' 
\delta 
    (
        \omega 
            - E
            - E' 
            + E_{\vb{k}, -\sigma } 
    )
\\& \times 
\left[
\text{sgn} \left(
    -\omega - E_{\vb{k}\sigma} - 2E_{\vb{k}, -\sigma} 
    + 2E
    + 2E'
    - u_{\vb{k}}
\right)
    \frac{
    (1 - n_{\vb{k}, -\sigma } ) 
        (1 - n_{\vb{k}\sigma } )
    \theta(-E')
    \theta(-E)
    }{
        |- E_{\vb{k}, -\sigma } 
        + E'
        + E
        - E_{\vb{k}\sigma}
        - u_{\vb{k}}|
    }
\right. \\& \left. 
+
\text{sgn} \left(
    -\omega - E_{\vb{k}\sigma} - 2E_{\vb{k}, -\sigma} 
    + 2E
    + 2E'
    + u_{\vb{k}}
\right)
    \frac{
    n_{\vb{k}, -\sigma } 
        n_{\vb{k}\sigma }
    \theta(E)
    \theta(E')
    }{
        |- E_{\vb{k}, -\sigma } 
        + E'
        + E
        - E_{\vb{k}\sigma}
        + u_{\vb{k}}|
    }
\right]
. 
\end{split}
\end{equation}
We then perform the integrals over $E$ and $E'$:  
\begin{equation}
\begin{split}
\text{Im}
&
\Sigma_{\sigma}^{\text{\ref{fig:_se_c}}} (\omega, \vb{k})
\approx
\\&
-\pi V^2 D^2 
\left[
- 
\text{sgn}\left(
    \omega - E_{\vb{k}\sigma } - u_{\vb{k}}
\right)
    \frac{
        (1 - n_{\vb{k}, -\sigma } ) 
        (1 - n_{\vb{k}\sigma } )
    }{
        |\omega - E_{\vb{k}\sigma } - u_{\vb{k} }|
    }
    + 
\text{sgn}\left(
    \omega - E_{\vb{k}\sigma } + u_{\vb{k}}
\right)
    \frac{
        n_{\vb{k}, -\sigma } 
        n_{\vb{k}\sigma }
    }{
        |\omega - E_{\vb{k}\sigma } + u_{\vb{k}}|
    }
\right]
u_{\vb{k} } (E_{\vb{k}, -\sigma } + \omega ) 
\label{eq:app_c_final_omega_simp}
. 
\end{split}
\end{equation}
As $u_{\vb{k}}$ approaches to 0,  the quantity (\ref{eq:app_c_final_omega_simp}) vanishes. We let $\omega \rightarrow E_{\vb{k}\sigma}$. The sign functions become $\text{sgn}(-u_{\vb{k}})$ and $\text{sgn}(u_{\vb{k}})$. Hence, (\ref{eq:app_c_final_omega_simp}) becomes 
\begin{align}
\Sigma_{\sigma}^{\text{\ref{fig:_se_c}}} (\omega, \vb{k})
\approx
&
-\pi V^2 D^2 
\left[
    - 
    \text{sgn} (-u_{\vb{k}})
    \frac{
        (1 - n_{\vb{k}, -\sigma } ) 
        (1 - n_{\vb{k}\sigma } )
    }{
         |-u_{\vb{k} }|
    }
    + 
    \text{sgn} (u_{\vb{k}})
    \frac{
         n_{\vb{k}, -\sigma } 
        n_{\vb{k}\sigma }
    }{
         |u_{\vb{k}}|
    }
\right]
u_{\vb{k} } (E_{\vb{k}, \sigma } + E_{\vb{k}, -\sigma } ) 
\\=&
-2\pi V^2 D^2 \frac{u_{\vb{k}}}{|u_{\vb{k}}|}
\text{sgn}(u_{\vb{k}})
\left[
    \epsilon_{\vb{k}} 
    (1 - n_{\vb{k}, -\sigma } ) 
        (1 - n_{\vb{k}\sigma } )
    +
    (\epsilon_{\vb{k}}+u_{\vb{k}}) 
     n_{\vb{k}, -\sigma } 
        n_{\vb{k}\sigma }
\right]
\\=& 
-2\pi V^2 D^2 
\left[
    \epsilon_{\vb{k}} 
    (1 - n_{\vb{k}, -\sigma } ) 
        (1 - n_{\vb{k}\sigma } )
    +
    (\epsilon_{\vb{k}}+u_{\vb{k}}) 
     n_{\vb{k}, -\sigma } 
        n_{\vb{k}\sigma }
\right]
\label{eq:app_c_omegaE}
, 
\end{align}
where the value of $E_{\vb{k}\sigma }$ and $E_{\vb{k}, -\sigma}$ depend on corresponding occupation number: for $n_{\vb{k}\sigma}=n_{\vb{k}, -\sigma} = 0$, $E_{\vb{k}\sigma } = E_{\vb{k}, -\sigma} = \epsilon_{\vb{k}}$; with $n_{\vb{k}\sigma}=n_{\vb{k}, -\sigma} = 1$, $E_{\vb{k}\sigma } = E_{\vb{k}, -\sigma} = \epsilon_{\vb{k}} + u_{\vb{k}}$. We examine the conditions with two different occupations: as $n_{\vb{k}\sigma } = n_{\vb{k}, -\sigma } = 0$, 
\begin{align}
\text{Im }\Sigma_{\sigma}^{\text{\ref{fig:_se_c}}} (\vb{k})
= 
\text{Im }\Sigma_{\sigma}^{\text{\ref{fig:_se_d}}} (\vb{k})
\approx
-2\pi V^2 D^2 
    \epsilon_{\vb{k}} 
. 
\label{eq:cde_0}
\end{align}
According to Eq. (\ref{eq:boundary2}), when we approach both the Fermi-arcs and the pseudo-Fermi surfaces from $n_{\vb{k}} = 0 $ region, $\epsilon_{\vb{k}} \rightarrow 0 $ and thus (\ref{eq:cde_0}) vanishes. As $n_{\vb{k}\sigma } = n_{\vb{k}, -\sigma } = 2$, 
\begin{align}
\text{Im }\Sigma_{\sigma}^{\text{\ref{fig:_se_c}}} (\vb{k})
=
\text{Im }\Sigma_{\sigma}^{\text{\ref{fig:_se_d}}} (\vb{k})
\approx
-2\pi V^2 D^2 
    (\epsilon_{\vb{k}}+u_{\vb{k}}) 
. 
\label{eq:cde_0}
\end{align}
According to Eq. (\ref{eq:boundary2}) and Eq. (\ref{eq:boundary1}), when we approach the Fermi-arcs from $n_{\vb{k}} = 2 $ region, $\epsilon_{\vb{k}} + u_{\vb{k}} \rightarrow 0 $, and when approaching to the pseudo-Fermi surfaces, we have $\epsilon_{\vb{k}} + u_{\vb{k}} \rightarrow u_{\vb{k}}$. Hence, we conclude that overall the result of (\ref{eq:app_c_omegaE}) is zero or linear in $u_{\vb{k}}$.

We then consider the self-energy diagram in Fig. \ref{fig:_se_g}. 
\begin{equation}
\begin{split}	
\Sigma_{\sigma}^{\text{\ref{fig:_se_g}}} (\omega, \vb{k})
=&
    i^3 V^2
    \int \frac{d^2 q}{(2\pi)^2 } 
    \int \frac{ds^0}{2\pi} \frac{dq^0}{2\pi} \frac{dk'^0}{2\pi} 
    G_\sigma^0 (s^0, \vb{k}) 
    G_\sigma^0 (s^0-q^0, \vb{k}-\vb{q})
    G_{-\sigma}^0 ({k'}^0, \vb{k}) 
    \\ & \times
    G_{-\sigma}^0 ({k'}^0+\omega-s^0, \vb{k})
    G_{-\sigma}^0 ({k'}^0-q^0, \vb{k}'-\vb{q}) 
    \Gamma_{\text{ph}} (s^0-{k'}^0, \vb{k})
    . 
\end{split}
\end{equation}
The frequency integral gives 
\begin{equation}
\begin{split}
\Sigma_{\sigma}^{\text{\ref{fig:_se_g}}} (\omega, \vb{k})
=& 
i^6 V^2 u_{\vb{k} }
\int \frac{d^2 q}{(2\pi)^2}
\\& \times
\left[
  \frac{
         n_{\vb{k}, -\sigma }
        ( 1 - n_{\vb{k}\sigma } ) 
        ( 1 - n_{\vb{k}-\vb{q}, -\sigma } )
        n_{\vb{k}-\vb{q}, \sigma }
    }{
        (
            \omega 
            + E_{\vb{k}-\vb{q}, -\sigma } 
            - E_{\vb{k}-\vb{q}, \sigma } 
            - E_{\vb{k}, -\sigma } 
            - i\eta 
        )
        (
            E_{\vb{k}, -\sigma } 
            - E_{\vb{k}-\vb{q}, -\sigma }
            + E_{\vb{k}-\vb{q}, \sigma }
            - E_{\vb{k}\sigma}
            - u_{\vb{k}}
            + i\eta 
        )
    }
\right. \\& \left. \quad  
    +
    \frac{
        (1 - n_{\vb{k}, -\sigma } ) 
        n_{\vb{k}\sigma }
        n_{\vb{k}-\vb{q}, -\sigma }
        ( 1 - n_{\vb{k}-\vb{q}, \sigma } )
    }{
        (
            \omega 
            + E_{\vb{k}-\vb{q}, \sigma } 
            - E_{\vb{k}-\vb{q}, \sigma } 
            - E_{\vb{k}, -\sigma } 
            + i\eta 
        )
        (
            E_{\vb{k}, -\sigma } 
            - E_{\vb{k}-\vb{q}, -\sigma }
            + E_{\vb{k}-\vb{q}, \sigma }
            - E_{\vb{k}\sigma}
            + u_{\vb{k}}
            - i\eta 
        )
    }
\right]
.
\end{split}
\end{equation}
The imaginary part is 
\begin{equation}
\begin{split}
\label{eq:se_d_img}
\text{Im }\Sigma_{\sigma}^{\text{\ref{fig:_se_g}}} (\omega, \vb{k})
= &
i^6 \pi V^2 
u_{\vb{k}}
\int \frac{d^2 q}{(2\pi)^2}
\delta 
    (
        \omega 
        + E_{\vb{k}-\vb{q}, -\sigma } 
        - E_{\vb{k}-\vb{q}, \sigma } 
        - E_{\vb{k}, -\sigma } 
    )
 \\& \times 
\left[
-
\text{sgn}\left(
    \omega + E_{\vb{k}\sigma }
    - 2E_{\vb{k}, -\sigma }
    + 2E_{\vb{k}-\vb{q}, -\sigma } - 2E_{\vb{k}-\vb{q}, \sigma }
    + u_{\vb{k}}
\right)
\frac{
    n_{\vb{k}, -\sigma }
        ( 1 - n_{\vb{k}\sigma } ) 
        ( 1 - n_{\vb{k}-\vb{q}, -\sigma } )
        n_{\vb{k}-\vb{q}, \sigma }
}{
        |E_{\vb{k}, -\sigma } 
            - E_{\vb{k}-\vb{q}, -\sigma }
            + E_{\vb{k}-\vb{q}, \sigma }
            - E_{\vb{k}\sigma}
            - u_{\vb{k}}|
}
\right. \\& \left. 
+
\text{sgn}\left(
    \omega + E_{\vb{k}\sigma }
    - 2E_{\vb{k}, -\sigma }
    + 2E_{\vb{k}-\vb{q}, -\sigma } - 2E_{\vb{k}-\vb{q}, \sigma }
    - u_{\vb{k}}
\right)
\frac{
        (1 - n_{\vb{k}, -\sigma } ) 
        n_{\vb{k}\sigma }
        n_{\vb{k}-\vb{q}, -\sigma }
        ( 1 - n_{\vb{k}-\vb{q}, \sigma } )
}{
        |E_{\vb{k}, -\sigma } 
        - E_{\vb{k}-\vb{q}, -\sigma }
        + E_{\vb{k}-\vb{q}, \sigma }
        - E_{\vb{k}\sigma}
        + u_{\vb{k}}|
}
\right] 
.
\end{split}
\end{equation}
After simplification we have 
\begin{equation}
\begin{split}
\label{eq:se_g_mom}
\text{Im}\Sigma_{\sigma}^{\text{\ref{fig:_se_g}}} (\omega, \vb{k})
= &
i^6 \pi V^2 
u_{\vb{k}}
\int \frac{d^2 q}{(2\pi)^2}
\\& \times 
\left[
-
\text{sgn}\left(
    \omega + E_{\vb{k}\sigma }
    - 2E_{\vb{k}, -\sigma }
    - 2u_{\vb{k}-\vb{q}}
    + u_{\vb{k}}
\right)
    \delta 
    (
        \omega 
        + u_{\vb{k}-\vb{q}}
        - E_{\vb{k}, -\sigma}
    )
\frac{
    n_{\vb{k}, -\sigma }
        ( 1 - n_{\vb{k}\sigma } )
        \theta(\epsilon_{\vb{k}-\vb{q}}+u_{\vb{k}-\vb{q}})
        \theta(-\epsilon_{\vb{k}-\vb{q}})
}{
        |E_{\vb{k}, -\sigma } 
            - u_{\vb{k}-\vb{q}}
            - E_{\vb{k}\sigma}
            - u_{\vb{k}}|
}
\right. \\& \left. 
+
\text{sgn}\left(
    \omega + E_{\vb{k}\sigma }
    - 2E_{\vb{k}, -\sigma }
    + 2u_{\vb{k}-\vb{q}}
    - u_{\vb{k}}
\right)
\delta 
    (
        \omega 
        - u_{\vb{k}-\vb{q}}
        - E_{\vb{k}, -\sigma}
    )
\frac{
        (1 - n_{\vb{k}, -\sigma } ) 
        n_{\vb{k}\sigma }
        \theta(\epsilon_{\vb{k}-\vb{q}}+u_{\vb{k}-\vb{q}})
        \theta(-\epsilon_{\vb{k}-\vb{q}})
}{
        |E_{\vb{k}, -\sigma } 
            + u_{\vb{k}-\vb{q}}
            - E_{\vb{k}\sigma}
            + u_{\vb{k}}|
}
\right] 
.
\end{split}
\end{equation}
We let $E_{\vb{q}'} = \epsilon_{\vb{k}-\vb{q}}+u_{\vb{k}-\vb{q}}$, and we transform $\int qdq q'dq' \rightarrow (2\pi D)^2 \int dEdE' $ where $\epsilon_{\vb{k}-\vb{q}} = \epsilon_{\vb{k}q\phi} \rightarrow E$ and $E_{\vb{q}'}=E_{q'\phi'}\rightarrow E'$. After we ignore the quantities of the order $O(1)$, Eq. (\ref{eq:se_g_mom}) becomes  
\begin{equation}
\begin{split}
\text{Im}\Sigma_{\sigma}^{\text{\ref{fig:_se_g}}} (\omega, \vb{k})
\approx &
i^6 \frac{\pi}{(2\pi)^2} (2\pi)^2 D^2 V^2 u_{\vb{k}} 
\int dE dE' 
\\& \times 
\left[
-
\text{sgn}\left(
    \omega + E_{\vb{k}\sigma} - 2E_{\vb{k}, -\sigma}
    - 2E' + 2E + u_{\vb{k}}
\right)
\delta 
    (
        \omega 
        + E' - E
        - E_{\vb{k}, -\sigma}
    )
\frac{
    n_{\vb{k}, -\sigma }
        ( 1 - n_{\vb{k}\sigma } )
        \theta(E')
        \theta(-E)
}{
        |E_{\vb{k}, -\sigma } 
            - E' + E
            - E_{\vb{k}\sigma}
            - u_{\vb{k}}|
}
\right. \\& \left. 
+  
\text{sgn}\left(
    \omega + E_{\vb{k}\sigma} - 2E_{\vb{k}, -\sigma}
    - 2E' + 2E - u_{\vb{k}}
\right)
\delta(
        \omega 
        - E' + E
        - E_{\vb{k}, -\sigma}
    )
\frac{
        (1 - n_{\vb{k}, -\sigma } ) 
        n_{\vb{k}\sigma }
        \theta( E')
        \theta( -E )
}{
        |E_{\vb{k}, -\sigma } 
            + E' - E
            - E_{\vb{k}\sigma}
            + u_{\vb{k}}|
}
\right] 
,
\end{split}
\end{equation}
where $D$ is density of states. This gives
\begin{align}
\text{Im}\Sigma_{\sigma}^{\text{\ref{fig:_se_g}}} (\omega, \vb{k})
\approx &
i^6 \pi V^2 D^2 u_{\vb{k}} 
\left[
-
    \text{sgn}\left(
    3\omega + E_{\vb{k}\sigma } - 4E_{\vb{k}, -\sigma }
    + u_{\vb{k}}
    \right)
    \frac{
        n_{\vb{k},-\sigma }(1-n_{\vb{k}\sigma })
        (-\omega + E_{\vb{k}, -\sigma })
    }{
        |\omega - E_{\vb{k}\sigma } - u_{\vb{k}}|
        }
    \right. \\& \left. 
+
    \text{sgn}\left(
    3\omega + E_{\vb{k}\sigma } - 4E_{\vb{k}, -\sigma }
    - u_{\vb{k}}
    \right)
    \frac{
        (1-n_{\vb{k},-\sigma })n_{\vb{k}\sigma }
        (\omega - E_{\vb{k}, -\sigma })
    }{
        |\omega - E_{\vb{k}\sigma } + u_{\vb{k}}|
    }
\right]
. 
\end{align}
We let $\omega \rightarrow E_{\vb{k}\sigma}$, 
\begin{align}
\text{Im}\Sigma_{\sigma}^{\text{\ref{fig:_se_g}}} (\vb{k})
\approx &
- \pi V^2 D^2 u_{\vb{k}}
(n_{\vb{k},-\sigma} - n_{\vb{k}\sigma } )^2 
. 
\end{align}

\begin{figure}[!h]
    \centering
\begin{subfigure}[t]{0.45\textwidth} 
    \centering
    \includegraphics[width=7.8cm]{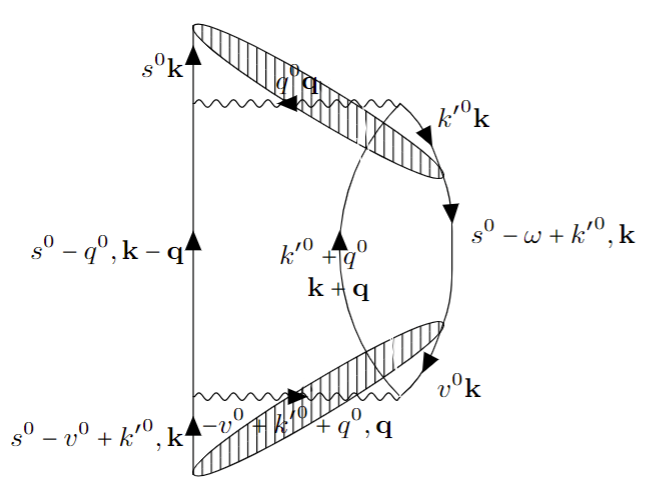}
    \caption*{Fig. \ref{fig:_se_e}}
    \label{fig:se_e}
\end{subfigure}
\begin{subfigure}[t]{0.45\textwidth} 
    \centering
    \includegraphics[width=7.5cm]{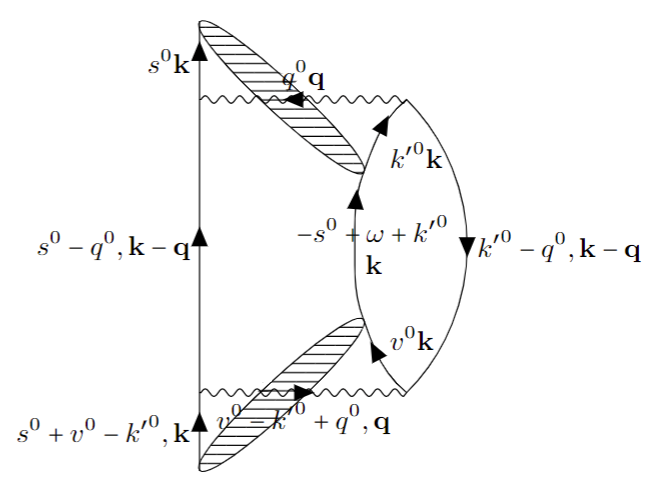}
    \caption*{Fig. \ref{fig:_se_i}}
    \label{fig:se_i}
\end{subfigure}
    \label{fig:se2}
\end{figure}

Finally we consider the self-energy diagrams in Fig. \ref{fig:_se_e} and Fig. \ref{fig:_se_i}. The expression for Fig. \ref{fig:_se_e} is 
\begin{equation}	
\begin{split}
& 
\Sigma_{\sigma}^{\text{\ref{fig:_se_e}}} (\omega, \vb{k})
=
    i^3 V^2
    \int \frac{d^2 q}{(2\pi)^2 } 
    \int \frac{ds^0}{2\pi} \frac{dq^0}{2\pi} \frac{dk'^0}{2\pi}
    G_\sigma^0 (s^0, \vb{k}) 
    G_\sigma^0 (s^0-q^0, \vb{k}-\vb{q})
    G_\sigma^0 (s^0-v^0+{k'}^0, \vb{k} )
    G_{-\sigma}^0 ({k'}^0, \vb{k}) 
    G_{-\sigma}^0 (v^0, \vb{k})
    \\& \times \quad 
    G_{-\sigma}^0 (s^0-\omega+{k'}^0, \vb{k})
    G_{-\sigma}^0 ({k'}^0+q^0, \vb{k}+\vb{q}) 
    (\Gamma_{\text{pp}} (s^0+{k'}^0, \vb{k}))^2 
. 
\end{split}
\end{equation}
The frequency integral gives 
\begin{equation}	
\begin{split}
i^6 V^2 u_{\vb{k} }^2 
\int \frac{d^2 q}{(2\pi)^2}
&
\left[
     \frac{
        (1 - n_{\vb{k}, -\sigma } ) 
        (1 - n_{\vb{k}\sigma } )
        n_{\vb{k}+ \vb{q}, -\sigma }
        n_{\vb{k}-\vb{q}, \sigma } 
    }{
        (
            - \omega 
            + E_{\vb{k}-\vb{q}, -\sigma } 
            + E_{\vb{k}+\vb{q}, -\sigma } 
            - E_{\vb{k}, -\sigma } 
            + i\eta 
        )
        (
            - E_{\vb{k}\sigma} 
            + E_{\vb{k}-\vb{q}, -\sigma } 
            + E_{\vb{k}+\vb{q}, -\sigma } 
            - E_{\vb{k}, -\sigma } 
            - u_{\vb{k}}
            + i\eta 
        )^2 
    }
\right. \\& \left. \quad +
    \frac{
        n_{\vb{k}, -\sigma } 
        n_{\vb{k}\sigma }
        (1 - n_{\vb{k}+ \vb{q}, -\sigma })
        (1 - n_{\vb{k}-\vb{q}, \sigma } )
    }{
        (
            - \omega 
            + E_{\vb{k}-\vb{q}, -\sigma } 
            + E_{\vb{k}+\vb{q}, -\sigma } 
            - E_{\vb{k}, -\sigma } 
            - i\eta 
        )
        (
            - E_{\vb{k}\sigma} 
            + E_{\vb{k}-\vb{q}, -\sigma } 
            + E_{\vb{k}+\vb{q}, -\sigma } 
            - E_{\vb{k}, -\sigma }
            + u_{\vb{k}}
            - i\eta 
        )^2 
    }
\right]
. 
\end{split}
\end{equation}
The expression for Fig. \ref{fig:_se_i} is 
\begin{equation}	
\begin{split}
& 
\Sigma_{\sigma}^{\text{\ref{fig:_se_i}}} (\omega, \vb{k})
=
    i^3 V^2
    \int \frac{d^2 q}{(2\pi)^2 } 
    \int \frac{ds^0}{2\pi} \frac{dq^0}{2\pi} \frac{dk'^0}{2\pi}
    G_\sigma^0 (s^0, \vb{k}) 
    G_\sigma^0 (s^0-q^0, \vb{k}-\vb{q})
    G_\sigma^0 (s^0-v^0+{k'}^0, \vb{k} )
    G_{-\sigma}^0 ({k'}^0, \vb{k}) 
    G_{-\sigma}^0 (v^0, \vb{k})
    \\& \times \quad 
    G_{-\sigma}^0 (-s^0+\omega+{k'}^0, \vb{k})
    G_{-\sigma}^0 ({k'}^0-q^0, \vb{k}-\vb{q}) 
    (\Gamma_{\text{ph}} (s^0-{k'}^0, \vb{k}))^2 
.
\end{split}
\end{equation}
The frequency integral gives 
\begin{equation}	
\begin{split}
i^6 V^2 u_{\vb{k} }^2 
\int \frac{d^2 q}{(2\pi)^2}
&
\left[
-
    \frac{
         n_{\vb{k}, -\sigma }
        ( 1 - n_{\vb{k}\sigma } ) 
        ( 1 - n_{\vb{k}-\vb{q}, -\sigma } )
        n_{\vb{k}-\vb{q}, \sigma }
    }{
        (
            \omega 
            + E_{\vb{k}-\vb{q}, -\sigma } 
            - E_{\vb{k}-\vb{q}, \sigma } 
            - E_{\vb{k}, -\sigma } 
            - i\eta 
        )
        (
            E_{\vb{k}, -\sigma } 
            - E_{\vb{k}-\vb{q}, -\sigma }
            + E_{\vb{k}-\vb{q}, \sigma }
            - E_{\vb{k}\sigma}
            - u_{\vb{k}}
            + i\eta 
        )^2 
    }
\right. \\& \left. \quad +
    \frac{
        (1 - n_{\vb{k}, -\sigma } ) 
        n_{\vb{k}\sigma }
        n_{\vb{k}-\vb{q}, -\sigma }
        ( 1 - n_{\vb{k}-\vb{q}, \sigma } )
    }{
        (
            \omega 
            + E_{\vb{k}-\vb{q}, -\sigma } 
            - E_{\vb{k}-\vb{q}, \sigma } 
            - E_{\vb{k}, -\sigma } 
            + i\eta 
        )
        (
            E_{\vb{k}, -\sigma } 
            - E_{\vb{k}-\vb{q}, -\sigma }
            + E_{\vb{k}-\vb{q}, \sigma }
            - E_{\vb{k}\sigma}
            + u_{\vb{k}}
            - i\eta 
        )^2 
    }
\right]
. 
\end{split}
\end{equation}
The imaginary parts are 
\begin{equation}	
\begin{split}
\text{Im }\Sigma_{\sigma}^{\text{\ref{fig:_se_i}}} (\omega, \vb{k}) =
& 
i^6 V^2 
u_{\vb{k}}^2 
\int \frac{d^2 q}{(2\pi)^2}
\delta 
    (
        \omega 
            - E_{\vb{k}-\vb{q}, \sigma } 
            - E_{\vb{k}+\vb{q}, -\sigma } 
            + E_{\vb{k}, -\sigma } 
    )
\\& \times 
\left\{
-
\text{sgn}\left[
    (
        - E_{\vb{k}, -\sigma } 
        + E_{\vb{k}-\vb{q}, \sigma }
        + E_{\vb{k}+\vb{q}, -\sigma }
        - E_{\vb{k}\sigma }
        - u_{\vb{k}}
    )(
        -2\omega - E_{\vb{k}\sigma } - u_{\vb{k}}
        - 3 E_{\vb{k}, -\sigma }
        + 3 E_{\vb{k}-\vb{q}, \sigma }
        + 3 E_{\vb{k}+\vb{q}, -\sigma }
    )
\right]
\right. \\& \left. \times 
\frac{
(1 - n_{\vb{k}, -\sigma } ) 
        (1 - n_{\vb{k}\sigma } )
        n_{\vb{k}+ \vb{q}, -\sigma }
        n_{\vb{k}-\vb{q}, \sigma } 
}{
    \left(
    - E_{\vb{k}, -\sigma } 
    + E_{\vb{k}+\vb{q}, -\sigma }
    + E_{\vb{k}-\vb{q}, \sigma }
    - E_{\vb{k}\sigma}
    - u_{\vb{k}}
    \right)^2 
}
\right. \\& \left. 
+
\text{sgn}\left[
    (
        - E_{\vb{k}, -\sigma } 
        + E_{\vb{k}-\vb{q}, \sigma }
        + E_{\vb{k}+\vb{q}, -\sigma }
        - E_{\vb{k}\sigma }
        + u_{\vb{k}}
    )(
        -2\omega - E_{\vb{k}\sigma } + u_{\vb{k}}
        - 3 E_{\vb{k}, -\sigma }
        + 3 E_{\vb{k}-\vb{q}, \sigma }
        + 3 E_{\vb{k}+\vb{q}, -\sigma }
    )
\right]
\right. \\& \left. \times 
\frac{
    n_{\vb{k}, -\sigma } 
        n_{\vb{k}\sigma }
        (1 - n_{\vb{k}+ \vb{q}, -\sigma })
        (1 - n_{\vb{k}-\vb{q}, \sigma } )
}{
    \left(
    - E_{\vb{k}, -\sigma } 
    + E_{\vb{k}+\vb{q}, -\sigma }
    + E_{\vb{k}-\vb{q}, \sigma }
    - E_{\vb{k}\sigma}
    + u_{\vb{k}}
    \right)^2 
}
\right\}
. 
\end{split}
\end{equation}
and 
\begin{equation}	
\begin{split}
\text{Im }\Sigma_{\sigma}^{\text{\ref{fig:_se_e}}} (\omega, \vb{k}) 
=& 
i^6 V^2 
u_{\vb{k}}^2 
\int \frac{d^2 q}{(2\pi)^2}
\delta 
    (
        \omega 
        + E_{\vb{k}-\vb{q}, -\sigma } 
        - E_{\vb{k}-\vb{q}, \sigma } 
        - E_{\vb{k}, -\sigma } 
    )
\\& \times 
\left\{
\text{sgn}\left[
    (
        E_{\vb{k}, -\sigma } 
        - E_{\vb{k}-\vb{q}, - \sigma }
        + E_{\vb{k}-\vb{q}, \sigma }
        - E_{\vb{k}\sigma }
        - u_{\vb{k}}
    )(
        2\omega + E_{\vb{k}\sigma } - u_{\vb{k}}
        - 3 E_{\vb{k}, -\sigma }
        + 3 E_{\vb{k}-\vb{q}, -\sigma }
        - 3 E_{\vb{k}-\vb{q}, \sigma }
    )
\right]
\right. \\& \left. \times 
\frac{
    n_{\vb{k}, -\sigma }
        ( 1 - n_{\vb{k}\sigma } ) 
        ( 1 - n_{\vb{k}-\vb{q}, -\sigma } )
        n_{\vb{k}-\vb{q}, \sigma }
}{
        \left(
        E_{\vb{k}, -\sigma } 
            - E_{\vb{k}-\vb{q}, -\sigma }
            + E_{\vb{k}-\vb{q}, \sigma }
            - E_{\vb{k}\sigma}
            - u_{\vb{k}}
        \right)^2 
}
\right. \\& \left.  
+ 
\text{sgn}\left[
    (
        E_{\vb{k}, -\sigma } 
        - E_{\vb{k}-\vb{q}, - \sigma }
        + E_{\vb{k}-\vb{q}, \sigma }
        - E_{\vb{k}\sigma }
        + u_{\vb{k}}
    )(
        2\omega + E_{\vb{k}\sigma } + u_{\vb{k}}
        - 3 E_{\vb{k}, -\sigma }
        + 3 E_{\vb{k}-\vb{q}, -\sigma }
        - 3 E_{\vb{k}-\vb{q}, \sigma }
    )
\right]
\right. \\& \left. \times 
\frac{
        (1 - n_{\vb{k}, -\sigma } ) 
        n_{\vb{k}\sigma }
        n_{\vb{k}-\vb{q}, -\sigma }
        ( 1 - n_{\vb{k}-\vb{q}, \sigma } )
}{
    \left(
        E_{\vb{k}, -\sigma } 
        - E_{\vb{k}-\vb{q}, -\sigma }
        + E_{\vb{k}-\vb{q}, \sigma }
        - E_{\vb{k}\sigma}
        + u_{\vb{k}}
    \right)^2 
}
\right\}
,
\end{split}
\end{equation}
The substitution and the transformation on the integral variable are the same as we did for Fig. \ref{fig:_se_c} and Fig. \ref{fig:_se_g}. The final results are 
\begin{equation}
\begin{split}
\text{Im }\Sigma_{\sigma}^{\text{\ref{fig:_se_e}}} (\omega, \vb{k}) 
\approx &
i^6 V^2  D^2 
\left\{
    \text{sgn} [(\omega - E_{\vb{k}, -\sigma } - u_{\vb{k}})^2]
    \frac{
    -
        (1-n_{\vb{k},-\sigma} ) 
        (1-n_{\vb{k}\sigma }) 
    }{
        (
        \omega - E_{\vb{k}\sigma } - u_{\vb{k} }
        )^2 
    }
    +
    \text{sgn} [(\omega - E_{\vb{k}, -\sigma } + u_{\vb{k}})^2]
    \frac{
       n_{\vb{k},-\sigma} 
       n_{\vb{k}\sigma }
    }{
        (
        \omega - E_{\vb{k}\sigma } + u_{\vb{k}}
        )^2 
    }
\right\}
\\& \times 
u_{\vb{k} }^2 
(E_{\vb{k}, -\sigma } + \omega ) 
,
\end{split}
\end{equation}

\begin{equation}
\begin{split}
\text{Im }\Sigma_{\sigma}^{\text{\ref{fig:_se_i}}} (\omega, \vb{k}) 
\approx &
i^6 V^2  D^2 
\left\{
-
    \text{sgn} [(\omega - E_{\vb{k}, -\sigma } - u_{\vb{k}})^2]
    \frac{
       (1-n_{\vb{k}\sigma} ) 
        n_{\vb{k}, -\sigma } 
    }{
        \left(
        \omega - E_{\vb{k}\sigma } - u_{\vb{k} }
        \right)^2 
    }
    +
    \text{sgn} [(\omega - E_{\vb{k}, -\sigma } + u_{\vb{k}})^2]
    \frac{
       n_{\vb{k}\sigma} 
       (1- n_{\vb{k}, -\sigma } )
    }{
        \left(
        \omega - E_{\vb{k}\sigma } + u_{\vb{k}}
        \right)^2 
    }
\right\}
\\& \times 
u_{\vb{k} }^2 
(\omega - E_{\vb{k}, -\sigma }) 
.
\end{split}
\end{equation}
As $\omega \rightarrow E_{\vb{k}\sigma}$, 
\begin{align}	
& 
\text{Im }\Sigma_{\sigma}^{\text{\ref{fig:_se_e}}} (\omega, \vb{k})
\approx 
-2\pi V^2 D^2 
\left[
    -
    \epsilon_{\vb{k}} 
    (1 - n_{\vb{k}, -\sigma } ) 
        (1 - n_{\vb{k}\sigma } )
    +
    (\epsilon_{\vb{k}}+u_{\vb{k}}) 
     n_{\vb{k}, -\sigma } 
        n_{\vb{k}\sigma }
\right]
\label{eq:app_e_omegaE}
,  
\\&
\text{Im }\Sigma_{\sigma}^{\text{\ref{fig:_se_i}}} (\omega, \vb{k})
\approx 
-\pi V^2 D^2 u_{\vb{k}}
(n_{\vb{k},-\sigma} - n_{\vb{k}\sigma } )^2
\label{eq:app_i_omegaE}
. 
\end{align}
Here we notice that (\ref{eq:app_i_omegaE}) is linear in $u_{\vb{k}}$ and thus it vanishes as $u_{\vb{k}} \rightarrow 0$. As for the result (\ref{eq:app_e_omegaE}), we perform the same analysis as we did for Fig. \ref{fig:_se_c} and \ref{fig:_se_d}, and hence we conclude: as it approaches to Fermi-arcs from either $n_{\vb{k}} = 0 $ or $n_{\vb{k}}=2$ regions, and approaches to pseudo-Fermi surfaces from $n_{\vb{k}}=0$ region, it always vanishes; as it approaches to pseudo-Fermi surfaces from $n_{\vb{k}}=2$ region, it results in a quantity proportional to $u_{\vb{k}}$.

\end{widetext}


\begin{references}

\bibitem{Proust} For a recent review, see, {\em e.g.}, Cyril Proust and Louis Taillefer, The Remarkable Underlying Ground States of Cuprate Superconductors, Annual Review of Condensed Matter Physics Vol. 10:409-429 (2019).

\bibitem{Yoshida} See, {\em e.g.}, Teppei Yoshida, Makoto Hashimoto, Inna M. Vishik, Zhi-Xun Shen, Atsushi Fujimori, Pseudogap, Superconducting Gap, and Fermi Arc in High-Tc Cuprates Revealed by Angle-Resolved Photoemission Spectroscopy,
J. Phys. Soc. Jpn. {\bf 81}, 011006 (2012).

\bibitem{Book} See, {\em e.g.}, Steven M. Girvin and Kun Yang, {\it Modern Condensed Matter Physics}, ISBN: 9781107137394, Cambridge University Press, Cambridge (March 2019).
    
\bibitem{Yang21} Kun Yang, {Exactly solvable model of Fermi arcs and pseudogap},  Phys. Rev. B {\bf 103}, 024529 (2021).

\bibitem{hk} Yasuhiro Hatsugai, and Mahito Kohmoto, Exactly Solvable Model of Correlated Lattice Electrons in Any Dimensions, Journal of the Physical Society of Japan {\bf 61}, 2056 (1992).

\bibitem{Baskaran} Ganapathy Baskaran, An Exactly Solvable Fermion Model: Spinons, Holons and a non-Fermi Liquid Phase, Modern Physics Letters B {\bf 5}, 643 (1991).

\bibitem{FW} Alexander L. Fetter, and John Dirk Walecka, {\it Quantum theory of many-particle systems} (Courier Corporation, 2012). 

\bibitem{phillips20} Philip W. Phillips, Luke Yeo, and Edwin W. Huang. Exact theory for superconductivity in a doped Mott insulator. Nature Physics 16, {\bf 12}: 1175-1180 (2020).

\bibitem{nesselrodt} R. D. Nesselrodt and J. K. Freericks. Exact solution of two simple non-equilibrium electron-phonon and electron-electron coupled systems: The atomic limit of the Holstein-Hubbard model and the generalized Hatsugai-Komoto model. Physical Review B 104, {\bf 15}: 155104 (2021).

\bibitem{li} Yu Li, Vivek Mishra, Yi Zhou, and Fu-Chun Zhang. Two-stage superconductivity in the Hatsugai-Kohomoto-BCS model, New Journal of Physics (2022).


\bibitem{zhong} Yin Zhong. Solvable periodic Anderson model with infinite-range Hatsugai-Kohmoto interaction: Ground-states and beyond. Physical Review B 106, {\bf 15}: 155119 (2022).



\end{references}
\end{document}